\documentclass{jfm}
\usepackage{graphicx}
\usepackage{bm}
\usepackage{epstopdf, epsfig}
\usepackage{caption}
\usepackage{subcaption}

\shorttitle{Settling behaviour of thin curved particles in quiescent fluid and turbulence}

\title{Settling behaviour of thin curved particles in quiescent fluid and turbulence}

\author{Timothy T. K. Chan\aff{1,2}, Luis Blay Esteban\aff{2}, Sander G. Huisman\aff{1}, John S. Shrimpton\aff{2} \and Bharathram Ganapathisubramani\aff{2} \corresp{\email{G.Bharath@soton.ac.uk}},}
\affiliation{ \aff{1}Physics of Fluids Group, Max Planck Center Twente for Complex Fluid Dynamics, Faculty of Science and Technology, MESA+ Research Institute, and J. M. Burgers Centre for Fluids Dynamics, University of Twente, P.O. Box 217, 7500 AE Enschede, The Netherlands
	\aff{2}Aerodynamics and Flight Mechanics Group, Faculty of Engineering and Physical Sciences, University of Southampton, Hampshire, SO17 1BJ, UK}

\begin{document}
	
	\maketitle
	
	\begin{abstract}	
		The motion of thin curved falling particles is ubiquitous in both nature and industry but is not yet widely examined. Here, we describe an experimental study on the dynamics of thin cylindrical shells resembling broken bottle fragments settling through quiescent fluid and homogeneous anisotropic turbulence. The particles have Archimedes numbers based on the mean descent velocity $0.75 \times 10^4 \lesssim Ar \lesssim 2.75 \times 10^4$. Turbulence reaching a Reynolds number of $Re_\lambda \approx 100$ is generated in a water tank using random jet arrays mounted in a co-planar configuration. After the flow becomes statistically stationary, a particle is released and its three-dimensional motion is recorded using two orthogonally positioned high-speed cameras. We propose a simple pendulum model that accurately captures the velocity fluctuations of the particles in still fluid and find that differences in the falling style might be explained by a {closer alignment between the particle's pitch angle and its velocity vector}. By comparing the trajectories under background turbulence with the quiescent fluid cases, we measure a decrease in the mean descent velocity in turbulence for the conditions tested. We also study the secondary motion of the particles and identify descent events that are unique to turbulence such as `long gliding' and `rapid rotation' events. Lastly, we show an increase in the radial dispersion of the particles under background turbulence and correlate the timescale of descent events with the local settling velocity.
	\end{abstract}
	
	\maketitle
	
	\section{Introduction}	\label{Sec.Intro}
	Solid particles settling through fluids are all around us. Some of these processes occur in natural environments, like falling leaves; while others happen in engineering processes or due to human activities. In fact, the latter often have detrimental effects on nature such as water and air pollution. Differences in the inertial characteristics of solid materials are also used in engineering applications to separate residues and reduce the human footprint on the environment. Standard and uniflow cyclones are extensively used to remove particulate matter (up to 10$\,$\textmu m) from the carrier fluid; e.g. remove sand and black powder in the natural gas industry \citep{Bahadori2014}, to improve clinker burning processes \citep{Wasilewski2017} and in solid--solid separation in the mineral processing industry \citep{Tripathy2015}. Hydrodynamic separators based on similar physical principles are also employed in the recycling industry \citep{esteban_three_2016}, where they classify materials based on the materials' inertial properties through interaction with turbulence. In this type of {device}, co-mingled waste is introduced into a container where background turbulence prevents plastics from sinking. In contrast, glass particles which struggle to follow vortical structures drop to the bottom of the tank, where a strong mean flow carries them to the next stage for further treatment. In these facilities, different turbulent regimes are found at various depths of the separator. Plastic-glass separation predominantly occurs in the middle region of the tank, where particle concentration is low and the turbulence is not modified by the solids. However, to improve the separation efficiency of these devices, a thorough understanding of settling characteristics of irregular particles in turbulence is required.
	
	Much research has been conducted on axisymmetric solids settling in quiescent fluid (see  \citealp{ern_wake-induced_2012} for a detailed review), and it is well accepted that particle dynamics are determined by three dimensionless numbers. These are: 1) the Reynolds number $Re = \langle V_z \rangle D/\nu$, where $\langle V_z \rangle$ stands for the particle mean descent velocity, $D$ for its characteristic lengthscale and $\nu$ for the fluid kinematic viscosity; 2) the dimensionless rotational inertia $I^*$, defined as the ratio of the moment of inertia of the particle over that of its solid of revolution with the same density as the fluid; and 3) the particle aspect ratio $D/h$, where $h$ denotes the object's thickness.
	
\begin{figure}
		\centering
		\includegraphics[]{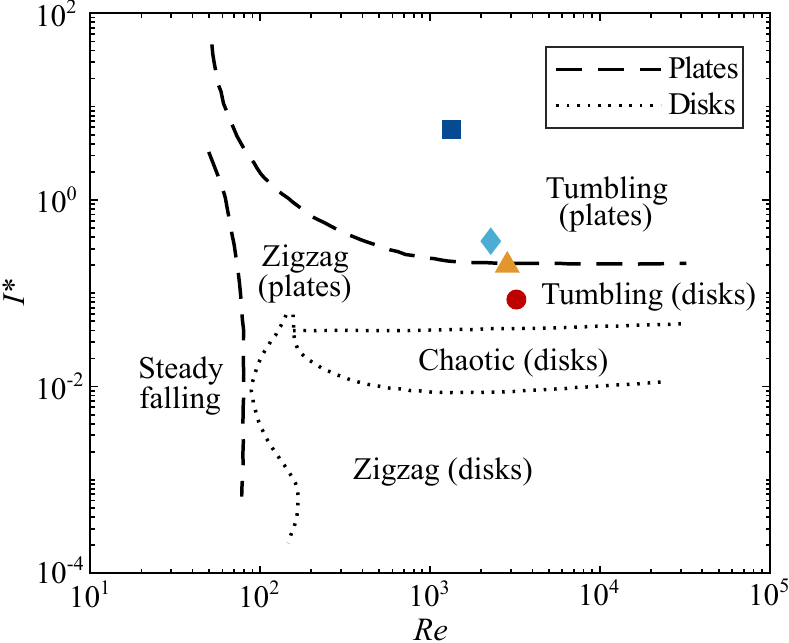}
		\caption{The $Re$--$I^*$ phase space explored in the current study. The regime boundaries are taken from \cite{field_chaotic_1997} and \cite{smith_autorotating_1971}. Markers denote the particles considered, whose properties are listed in table \ref{Table.ParticleProp}.}
		\label{Fig.ReI*Space}
	\end{figure}

	The most widely studied non-spherical particles are planar disks and rectangular plates \citep{stringham_behaviour_1969, field_chaotic_1997, ern_wake-induced_2012, auguste_falling_2013, smith_autorotating_1971, heisinger_coins_2014, mahadevan_tumbling_1999, zhong_experimental_2011, zhong_experimental_2013, lee_experimental_2013, chrust_numerical_2013}, whose falling styles share the same dominant features. Still, specific dynamics occur when the particle perimeter contains sharp edges \citep{esteban_edge_2018, esteban_study_2019, esteban_three_2019}. The four dominant regimes in both disks and rectangular plates are `steady fall', `zig-zag motion', `chaotic motion' and `tumbling motion'; and these are shown in the $Re-I^*$ phase space in figure \ref{Fig.ReI*Space}. When $Re$ is sufficiently small, a particle descends following a `steady fall' independent of its dimensionless moment of inertia. Under this mode, the solid falls vertically with oscillation amplitudes much smaller than its characteristic lengthscale. As $Re$ increases, the swaying motion grows and the particle transits into a `zig-zag motion' caused by vortex shedding.  Various types of zig-zag motions have been identified, ranging from `planar zig-zag' to more three-dimensional ones such as `spiralling' and `hula-hoop' motion \citep{auguste_falling_2013, zhong_experimental_2011}. From this point, as $I^*$ rises, the pitching motion of the particle overcomes the fluid torque damping it and the descent enters a `chaotic regime' where the particle flips over intermittently while exhibiting a zig-zag motion. As $I^*$ increases further, tumbling becomes more persistent and eventually continuous in the `tumbling motion' regime. Markers in figure \ref{Fig.ReI*Space} locate the solids investigated in this study in the $Re-I^*$ phase space originally determined for disks and plates \citep{willmarth_steady_1964, stringham_behaviour_1969, smith_autorotating_1971, field_chaotic_1997}. Details on defining the dimensionless numbers of these particles are included in \S \ref{Sec.Quiescent}.
	
	Regarding three-dimensional particles with curvature, spheroids, spheres, and cylinders are the canonical geometries that have been investigated in more detail. Oblate spheroids have the same principal falling styles as disks. However, as they become more spherical, zig-zag and chaotic descents vanish. Yet, when they are close to spheres, chaotic motion returns \citep{zhou_path_2017}. The dynamics of spheres is also very complex, with steady fall, oblique descent, horizontal oscillations due to vortex shedding; and helical and chaotic motions all observed \citep{jenny_instabilities_2004, veldhuis_experimental_2007, horowitz_effect_2010, zhou_chaotic_2015, ardekani_numerical_2016}. Fibre-like shapes such as prolate spheroids fall helically with no visible zig-zag motion \citep{ardekani_numerical_2016}. Still, as the aspect ratio increases and the particles become long cylinders, they settle rectilinearly or with oscillations along its axial direction \citep{horowitz_dynamics_2006, horowitz_vortex-induced_2010, toupoint_kinematics_2019}. We refer the reader to the comprehensive review by \cite{voth_anisotropic_2017} for the orientation of fibre-like particles under different flow conditions.
	
	Despite these studies, there has been little research on the kinematics of thin curved particles settling in quiescent fluid or under background turbulence. Nonetheless, this represents an interesting area of research not only for its fundamental significance but also for its industrial relevance.
	
	Literature concerning solids settling or rising in turbulence is far sparser due to the relative complexity of turbulence generation in a controlled environment. Studies generally focused on two issues: 1) the settling styles of individual particles and 2) how turbulence modifies the mean descent velocities. Note that research on the alignment or rotation of nearly buoyant solids with the carrier flow are not included. 
	
	Experiments generally focused on large particles so that their characteristic lengthscale lies within the range of turbulent inertial scales where solid--turbulence interactions are richer. Rising spheres in turbulence with a downward mean flow perform zig-zag motion or tumbling motion with the transition triggered by changes in $I^*$ \citep{mathai_flutter_2018}. Disks which undergo planar zig-zag motion in quiescent fluid settle differently in statistically stationary homogeneous anisotropic turbulence \citep{esteban_disks_2020}. The dominant features of the planar zig-zag mode in quiescent fluid are still observed. However, these are sometimes replaced by fast descents, tumbling events, long gliding sections, and hovering motions among others. The variety of descent scenarios demonstrate the complexity of the particle-turbulence interactions that occur during settling.	
	
	Despite the consensus that turbulence with zero mean flow changes the average settling velocity of spherical and non-spherical particles, a full understanding of this phenomenon has yet to be established. Four mechanisms that modify settling have been proposed to date: the {`preferential sweeping effect' \citep{maxey_gravitational_1986, maxey_gravitational_1987, tom_multiscale_2019},} nonlinear drag due to fluid acceleration \citep{ho_fall_1964}, `loitering effect' \citep{nielsen_turbulence_1993}, and vortex entrainment \citep{nielsen_motion_1984, nielsen_coastal_1992}. {Preferential sweeping} effect refers to the situation where particles are accelerated by the descending side of vortices as they spiral outwards from the core, whereas loitering effect simply means they stay relatively longer in upward flows. These four processes affect the local descend velocity $V_z$ differently, with the first increasing it and the others reducing it. In this framework, the settling rate modification depends on the relative importance of the competing mechanisms.
	
	The situation is further complicated as these effects may not be easily delineated, and opposite results regarding the descent speed have been reported. For droplets in isotropic turbulence, settling is enhanced when the ratio of the {particle's characteristic gravitational velocity} to the root-mean-square flow velocity fluctuations is smaller than unity and hindered otherwise \citep{good_settling_2014}. Nonlinear drag is proven to be vital for attenuating the descent in that case. On the other hand, simulations of finite rigid spheres in \cite{fornari_sedimentation_2016} found slower settling velocities in turbulence for all the {tested ratios of the mean descent velocity in quiescent fluid to the root-mean-square flow velocity fluctuations ($\langle V_q \rangle/u'_{rms}$).} However, the reduction in the mean descent velocity is greater when $\langle V_q \rangle/u'_{rms} < 1$ \citep{fornari_reduced_2016}. There, the authors attributed hindered settling to unsteady wake forces in addition to severe nonlinear drag due to horizontal oscillations. {Recently, \cite{tom_multiscale_2019} argued in the context of preferential sweeping that the parameter demarcating enhanced and hindered settling should account for the multiscale nature of particle--turbulence interactions. It is possible that the apparent contradictions can be reconciled with scale-dependent quantities, which have been employed to model pair statistics in turbulence \citep{bec_stochastic_2008}.}
	
	{The above results are restricted to spheres in turbulence.} Non-spherical solids with finite size and particle inertia add more complexity to the problem. Nearly neutrally buoyant cylinders of the order of the Taylor-microscale show small slip velocities in isotropic turbulence \citep{byron_slip_2019}, which may suggest nonlinear drag is not so important. Similarly, particles describing falling styles that reflect strong interactions with the media, where particle orientation plays a crucial role, also show an inconsistent behaviour with the velocity ratio proposed for small spherical particles. More specifically, disks falling in anisotropic turbulence where $\langle V_q \rangle/u'_{rms} > 1$ settle more rapidly than in quiescent fluid \citep{esteban_disks_2020}. Focusing on the frequency content of the trajectories, \cite{esteban_disks_2020} found that as turbulence intensity increases, the dominant frequency of the particles reduces; and this leads to enhanced settling. However, as different types of motions may occur in a single trajectory, the relation between the dominant frequency and the descent styles is not entirely clear.
	
	Given these contrasting results, it is obvious that a better understanding on how turbulence affects settling particles is needed, especially for complex geometries like non-spherical particles with curvature.
	
	We therefore study the kinematics of freely falling curved particles resembling bottle fragments. This paper is organised as follows. In \S \ref{Sec.Quiescent}, we present the experimental details of the quiescent fluid cases, discuss the results and propose a simple model for the motions observed. Next, we show the effects of background turbulence on the settling kinematics of the curved particles and discuss the results obtained in \S \ref{Sec.Turb}. Last, this paper concludes in \S \ref{Conclusion} with the main experimental findings and directions for future research.
	
	\section{Settling in quiescent fluid}	\label{Sec.Quiescent}
	\subsection{Methods}	\label{Sec.QuiescentMethods}
	To analyse the settling behaviour of thin curved objects, we drop bottle-fragment-like particles in a tank filled with tap water at room temperature (17$^\circ$C).
	
\begin{figure*}
		\centering
		\includegraphics[height=6cm]{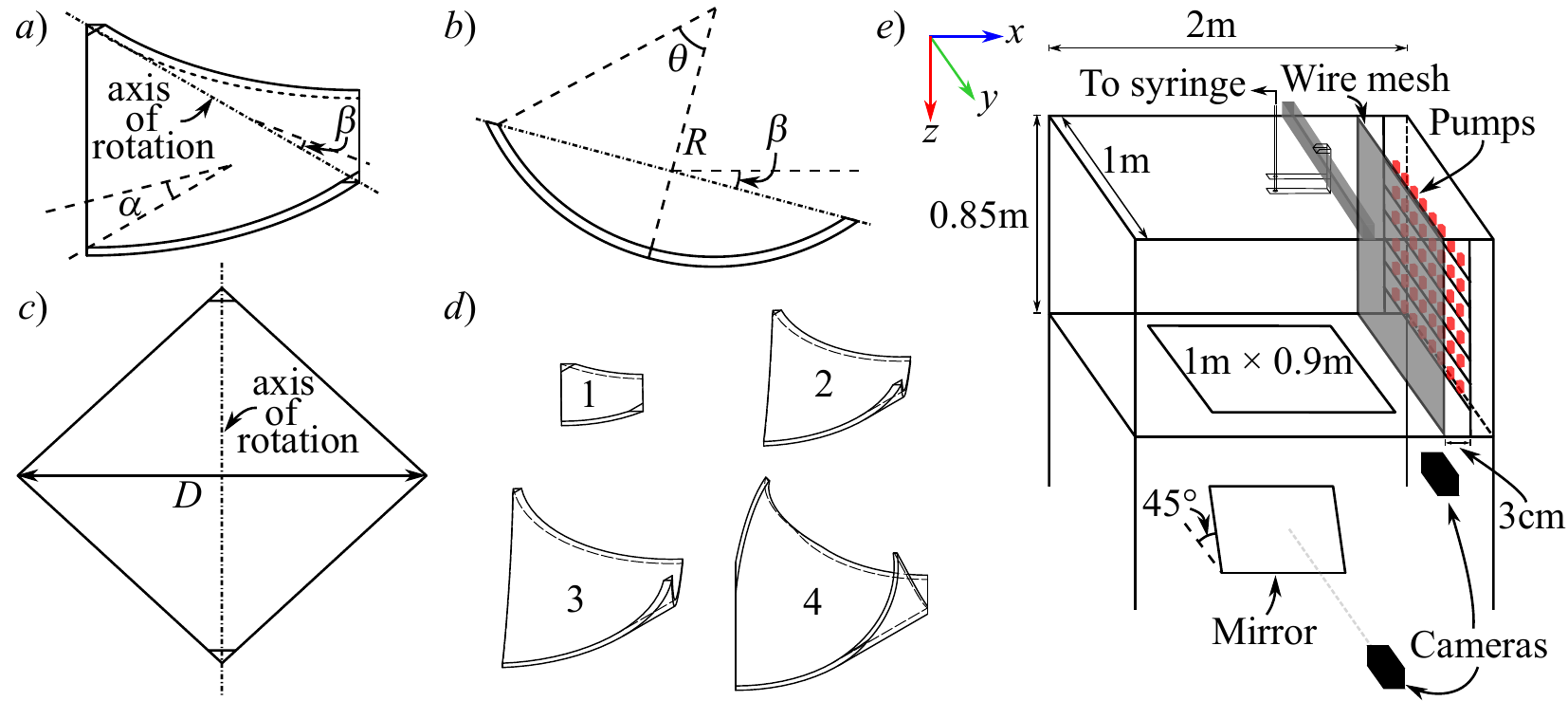}
		\caption{({\it a}) Bottle-fragment-like particle considered in this study. The dash-dot line is the axis of revolution used to obtain the dimensionless moment of inertia $I^*$, whereas $\alpha$ and $\beta$ are the pitch and roll angles respectively. ({\it b}) Front view, and ({\it c}) top view of particle with $\beta = 0$. ({\it d}) To-scale drawings of the four tested particles whose dimensions are listed in table \ref{Table.ParticleProp}. ({\it e}) Tank and release mechanism employed. Pumps and wire meshes are installed on both sides for symmetry, though only those on the right are shown to reduce clutter. The distance between the pumps is $165\,$cm. The tank rests on a steel frame with a rectangular window at the bottom to allow optical access. The coordinate system is shown on the top left.}
		\label{Fig.Setup}
	\end{figure*}

	Figure \ref{Fig.Setup}\,({\it a\/}) shows the geometry {used to model a broken cylindrical bottle. The particle has a parallelogramic projection and one non-zero principal curvature oriented along one of the diagonals of the parallelogram. Hence it is} completely defined by the radius of curvature of the original cylinder $R$, the subtended angle $\theta$, the diagonal length $D$, and the thickness $h$. {The values of these parameters are selected to mimic the dimensions of fragments processed in recycling plants \citep{esteban_three_2016}.} To delineate the effect of the different variables, $R$ is kept largely constant at approximately $19\,$mm and $\theta$ (thus $D$) is varied between {29}$^\circ$ and 115$^\circ$. The thickness also remains the same for all cases at $h = 1\,$mm, resulting in aspect ratios $D/h=22$ to $51$. It has been shown that the kinematics of freely falling disks at low $Re$ may differ even for very large aspect ratios near the `steady fall--zig-zag motion' transition \citep{auguste_falling_2013}. However, the $Re$ of the particles concerned are far from this boundary and small differences in the aspect ratio have little effect on their kinematics. Furthermore, thinner particles are not sufficiently rigid to withstand flow perturbations without deformation. We 3D-print all particles (Formlabs Form 2 printer) using a glass-reinforced rigid resin which results in a material with a flexural modulus $E \approx 3.7\,$GPa. A print resolution of $0.05\,$mm is used and the objects are wet sanded with P800 sandpaper for a smooth finish. Black spray paint, which amounts to less than 5\% of the particle mass, is applied to aid object detection. Table \ref{Table.ParticleProp} shows the particle dimensions determined post-production. The density ratios are also measured and found to be nearly constant across all the cases, with {$\rho^* = \rho_p/\rho_f = 1.70 \pm 0.05$. For the particle dimensionless moment of inertia, $I^* = I/I_0$, where $I$ is the object's moment of inertia and $I_0$ is the reference moment of inertia. Their precise definitions will be discussed below.
		
		Choosing a suitable $I^*$ is challenging without employing any assumptions regarding the particle behaviour, so past studies generally assume the particle concerned would mainly oscillate about a predetermined axis. Disks are supposed to rotate about its diameter. Presumably using this as an inspiration, for spheroids, \cite{zhou_path_2017} incorporated the ratio between the moment of inertia about the equatorial and polar axes in $I^*$ so the same axis of rotation is considered in the limit of disks. For $n$-sided polygons, \cite{esteban_edge_2018} adopted an analogous axis of rotation to disks when calculating the particle moment of inertia, but considered the perimeter of the particle relative to a circumscribed disk to correct for the characteristic length scale in the nondimensionalisation. To select the appropriate $I$ and $I_0$, we made an educated guess of the particle motion. Due to the presence of a dihedral, the particle should be more stable against rotations around its uncurved axis. We therefore expect it to rotate about the `axis of rotation' indicated in figure \ref{Fig.Setup}\,({\it a\/}). Therefore, in our case, $I$ is the object's moment of inertia about an axis passing through its centre of gravity and parallel to the line marked `axis of rotation' in figure \ref{Fig.Setup}\,({\it a\/}) and $I_0$ is the moment of inertia of a fluid-filled ellipsoid-like object generated by rotating the arc in figure \ref{Fig.Setup}\,({\it b\/}) about its vertices.}
	
	\begin{table}
		\centering
		\begin{tabular}{cccccc}
			Particle No. & Symbol & $R$ (mm) & $\theta$ ($^\circ$) & $D$ (mm) & $I^*$\\
			1 & \includegraphics[height = \fontcharht\font`\B]{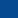} & 19 & 29 & 22 & 5.62\\
			2 & \includegraphics[height = \fontcharht\font`\R]{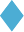} & 17 & 73 & 39 & 0.36\\
			3 & \includegraphics[height = \fontcharht\font`\B]{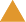} & 21 & 82 & 48 & 0.20\\
			4 & \includegraphics[height = \fontcharht\font`\B]{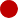} & 19 & 115 & 51 & 0.09
		\end{tabular}
		\caption{Dimensions of the particles dropped. $R$ is roughly constant at $19\,$mm while $\theta$ and $D$ are varied. {The uncertainty is captured by the number of digits reported.}}
		\label{Table.ParticleProp}
	\end{table}
	
	We release the particles in the glass tank shown in figure \ref{Fig.Setup}\,({\it e\/}). The tank, measuring $2\,$m $\times$ $1\,$m $\times$ $0.85\,$m, is mounted on a steel frame with a central $1\,$m $\times$ $0.9\,$m rectangular window at the bottom to enable optical access. In preparation for experiments in turbulence, the tank is equipped with an 8 $\times$ 6 bilge pump array (Rule 24, 360 GPH) on either side with a $5\,$mm squared wire mesh $3\,$cm downstream of the nozzles. The pumps are off for experiments in quiescent fluid, and the method of turbulence generation will be introduced later.
	
	To hold the particle, a pressure mechanism consisting of a syringe pump connected to a suction cup is used. First, the particle is affixed at $0$ pitch angle ($\alpha$, see figure \ref{Fig.Setup}\,{\it a\/}) to the suction cup by imposing a pressure deficit. Then, by slowly pushing the plunger of the syringe, the pressure is equalised to the atmosphere and the particle is released. Similar to the work by \cite{lau_progression_2018}, surface--particle interaction is minimised by adjusting the position of the suction cup to at least $1.5D$ below the water level and particle transient kinematics are discarded prior to the data post-processing. The object's surface is carefully verified to be bubble-free before release. Confinement effects are also negligible as the side walls of the tank remained at least 4$D$ from the object. A minimum of 8 minutes separate releases to allow any residual flow to dampen, and each particle is dropped {at least} 25 times to reduce random error.
	
	During each descent, the motion is recorded by two cameras operating at $60\,$Hz {using AF Nikkor 50\,mm objectives. While the top camera (JAI GO-5000M-USB; pixel size = 5\,\textmu m) captures the front view, the lower one (JAI GO-2400M-USB; pixel size = 5.86\,\textmu m)} records the bottom view through a mirror inclined 45$^\circ$. The camera aperture is set so that the contrast and the depth of field are sufficiently large for the entire descent; and the exposure times are adjusted accordingly. To ensure the three-dimensional particle motion reconstruction is accurate, the cameras are synchronized with a $5\,$V external signal (National Instruments USB-6212), aligned with respect to the tank by a vertical post and calibrated using a square grid. For the bottom camera, the {resolution at 5 different heights is calculated and a linear fit is used to obtain the image resolution as a function of depth. The resolution of both cameras is $\approx 0.2$\,mm/px, which corresponds to a magnification of $\approx 1/40$.}
	
	The three-dimensional position and orientation of the particle are extracted through Matlab. The image processing protocol to obtain the particle's centre of gravity is similar to the one proposed in  \cite{esteban_disks_2020}, where a background image is first subtracted from all frames. Then, a Gaussian filter with a {standard deviation} of $3\,${px} is applied and the resulting images are binarised before calculating the centres of gravity. Doing so, the script gives us the $(x,y)$ and $z$ coordinates of the object from the recordings of the  bottom and top cameras respectively.
	
	On the other hand, the pitch and roll angles of the particle, which are sketched in figure \ref{Fig.Setup}\,({\it a\/}), are evaluated by measuring the diagonal lengths in each frame. To calculate them, the corners of the particle are detected first in the binary image and then refined using the greyscale one. Finally, the position of one diagonal's midpoint relative to the other diagonal provides the signs of the pitch and roll angles. {The high-resolution image allows the pitch angle to be determined to $\lesssim 3^\circ$.} All the data have been smoothed by {Gaussian filters} to reduce high-frequency noise.
	
	In this study, we are interested in the non-transient particle kinematics. To remove the transient motions, we first construct the cumulative average of the instantaneous vertical velocity $\langle V_z \rangle_c$. By examining this magnitude, we observe that the particle descent velocity is stable after descending $2/3$ the tank depth ({$26D$ and $11D$} for the smallest and largest fragments respectively). Then, the cumulative average $\langle V_z \rangle_c$ at each vertical location is compared to the stabilised velocity, and the initial part of the trajectories where the deviation is greater than $\pm$10\% discarded. This threshold is robust, since {halving it to $\pm$5\% did not affect the results significantly}. Similarly, the last particle oscillation is ignored to eliminate motions affected by interactions with the bottom of the tank.
	
	\subsection{Results and discussion}
\begin{figure}
		\centering
		\includegraphics[]{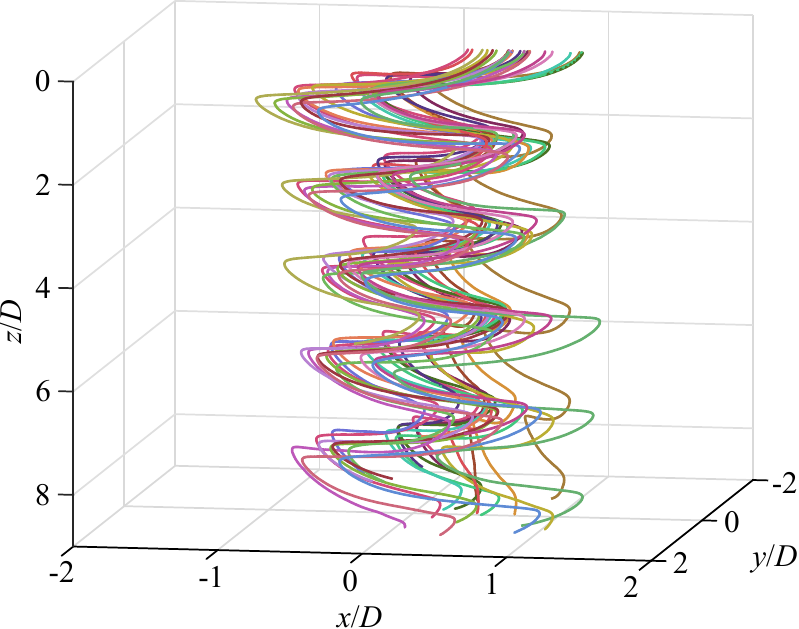}
		\caption{Reconstructed three-dimensional trajectories of particle No. 2 ($\theta = 73 ^\circ$, $D = 39\,$mm) in quiescent fluid.}
		\label{Fig.Q_TrajAll}
	\end{figure}

	Figure \ref{Fig.Q_TrajAll} shows the three-dimensional reconstruction of all 25 trajectories recorded for particle No. 2 in quiescent fluid after transient removal. All descents show periodic motions with a constant mean vertical velocity. However, the solid sometimes drifts horizontally in an apparently random direction as it settles. Similar trajectories are obtained for all types of particles tested. 
	
	\begin{table}
		\centering
		\begin{tabular}{ccccc}
			Particle No. & $V_{drift} $ (mm/s) & {$A_g/R_g$} & $\dot{\gamma}$ ($^\circ$/s) & $I_{roll}$ (g$\cdot$mm\textsuperscript{2})\\
			1 & $3.8\pm1.4$ & $0.05 \pm 0.04$ & $9.4 \pm 1.5$ & $8.7\times10^{-6}$\\
			2 & $3.2\pm1.4$ & $0.04 \pm 0.03$ & $7.4\pm0.9$ & $8.4\times10^{-5}$\\
			3 & $3.8 \pm 2.0$ & $0.02 \pm 0.02$ & $0.8\pm0.5$ & $2.0\times10^{-4}$\\
			4 & $2.5\pm1.1$ & $0.03\pm0.02$ & $2.8 \pm 0.7$ & $3.9\times10^{-4}$
		\end{tabular}
		\caption{The mean horizontal drift velocity $V_{drift}$, the {dimensionless gliding section amplitude $A_g/R_g$}, the precession rate $\dot{\gamma}$ of each particle, and the moment of inertia of rotating it about an axis passing through its centre of gravity and parallel to its uncurved diagonal $I_{roll}$.}
		\label{Table.DriftResults}
	\end{table}
	
	To ensure this motion can be neglected, we obtain the velocity associated with the horizontal drift $V_{drift}$ for all trajectories, see table \ref{Table.DriftResults}. The velocity magnitude appears to be insensitive to particle geometry and the horizontal drift has no obvious preferred direction. This suggests the drift is probably not inherent to the descent behaviour and may have originated from minute flows in the tank which are difficult to eliminate. This motion is unlikely to have been caused by the release mechanism since the flow induced by capillary waves decays exponentially in space. Experiments involving heavy cylinders in \cite{toupoint_kinematics_2019} also found similar behaviour and the authors argued this was related to large-scale fluid motions inside the tank.  For the subsequent analysis, the trajectories are dedrifted assuming $V_{drift}$ to be the average drift velocity over a square window centred about the current location and capturing one full period.
	
	\begin{figure*}
		\centering
		\includegraphics[]{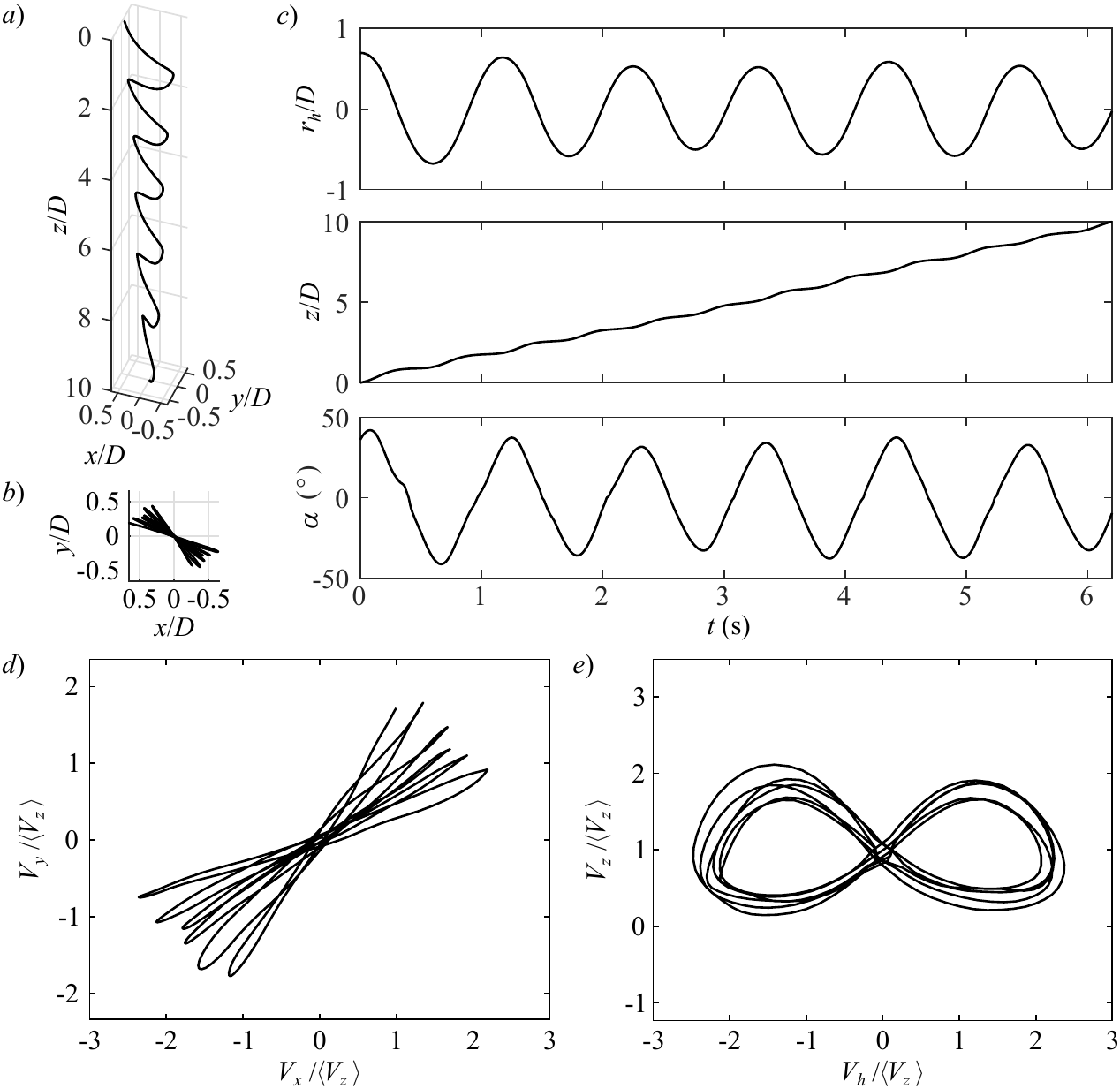}
		\caption{Descent of particle No. 2 ($\theta = 73 ^\circ$, $D = 39\,$mm) after removing transient motions and dedrifting. ({\it a}) Three-dimensional and ({\it b}) top view of the trajectory reconstruction. ({\it c}) (From top to bottom) The radial displacement along the direction of motion $r_h$, the depth $z$, and the pitch angle $\alpha$ plotted against time $t$. No rolling motions are observed. ({\it d}) Velocity in the $y$-direction $V_y$ plotted against that in the $x$-direction $V_x$. ({\it e}) Instantaneous vertical velocity $V_z$ against the horizontal velocity $V_h= \pm(V_x^2 +V_y^2)^{\frac{1}{2}}$ whose sign switches every swing. All velocities and positions are normalized with the mean descent velocity $\langle V_z \rangle$ and diagonal length $D$ of the particle respectively.}
		\label{Fig.Q_Th73OneTraj}
	\end{figure*}
	
	We then plot the settling behaviour of particle No. 2 in quiescent fluid in figure \ref{Fig.Q_Th73OneTraj} (see also the supplementary videos). As the object falls, it oscillates periodically in the $xy$-plane with a constant amplitude (figure \ref{Fig.Q_Th73OneTraj}\,{\it a--c\/}). At the beginning of each oscillation, the particle carries no horizontal velocity $V_h$ and shows a highly negative pitch angle $\alpha$ (pointing downwards). As the particle is not in equilibrium, it accelerates both downwards and horizontally along a direction inside its symmetry plane containing the uncurved diagonal until it reaches its maximum velocity, which occurs roughly at the middle of each swing. The particle then decelerates as $\alpha$ increases, drawing an arc-like trajectory. This process repeats itself in the opposite direction to complete one oscillation. In contrast to N-sided regular polygons \citep{esteban_three_2019}, the particles tested here always travel in a preferred orientation, that is: along the flat diagonal. Also, no rolling motions are detected which {agrees with our expectation in the discussion on $I^*$}.
	
	While it is obvious that the particles fall in a zig-zag fashion, whether the trajectories observed are planar or three-dimensional is not evident. Here, we use an analogous approach as the one proposed in \cite{esteban_edge_2018} where {each trajectory is split into `gliding' and `turning' sections by local extrema of the instantaneous descent velocity. The amplitude of each gliding section $A_g$, defined as half the planar displacement of the gliding section, is compared to its radius of curvature $R_g$} in the top-down view (figure \ref{Fig.Q_Th73OneTraj}\,{\it b\/}). All trajectories tested satisfy the criterion {$A_g/R_g < 0.1$} (table \ref{Table.DriftResults}), and therefore are considered to be within the `planar zig-zag' mode.
	
	Other features of `planar zig-zag' trajectories are also observed: oscillations in the $z$-direction have twice the frequency of those in the horizontal (figure \ref{Fig.Q_Th73OneTraj}\,{\it c\/}) \citep{zhong_experimental_2013}, and the velocity phase plot describes a characteristic butterfly shape (figure \ref{Fig.Q_Th73OneTraj}\,{\it e\/}) \citep{auguste_falling_2013}. Despite these similarities, the motion of bottle-fragment particles differ from disks in the sense that disks yaw almost 180$^\circ$ at every horizontal extremum \citep{zhong_experimental_2013}, but this does not occur for the particles tested.
	
	While certain disks exhibit three-dimensional `hula-hoop' descents which precess \citep{auguste_falling_2013}, and figure \ref{Fig.Q_Th73OneTraj}\,({\it d\/}) somewhat resembles such a mode, it is clear that the particles concerned do not fall this way. This is because `hula-hoop' settling has an ellipsoidal profile of $V_y$ against $V_x$. Instead, the precession observed here probably emerges due to {another reason}.
	
	To examine this feature, {we further studied the gliding and turning sections.} As negligible rotation occurs in the gliding sections, they are approximated by straight lines in the $xy$-plane. Therefore, {rotations have to occur during the turning sections} and the precession rate $\dot{\gamma}$ can be defined as the rate at which the gliding sections rotate, see table \ref{Table.DriftResults}. We observe that $\dot{\gamma}$ decreases as the rotational inertia about an axis parallel to the flat diagonal $I_{roll}$ increases. Thus, we hypothesize that tiny fluid fluctuations due to {residual flows} can explain the precession. These fluctuations may imperceptibly cause the object to roll hence precess in the turning sections.
	
\begin{figure}
		\centering
		\includegraphics[]{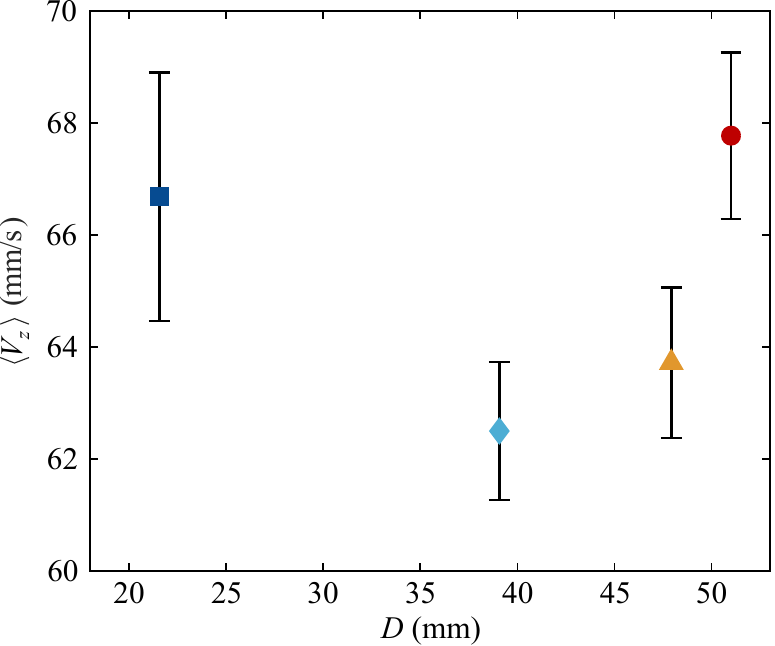}
		\caption{The mean settling velocities $\langle V_z \rangle$ of the particles. Unless further specified, the definitions of the data markers follow table \ref{Table.ParticleProp} and vertical error bars represent the standard deviation of the measurements.}
		\label{Fig.Q_meanUz}
	\end{figure}

\begin{figure}
		\centering
		\includegraphics[]{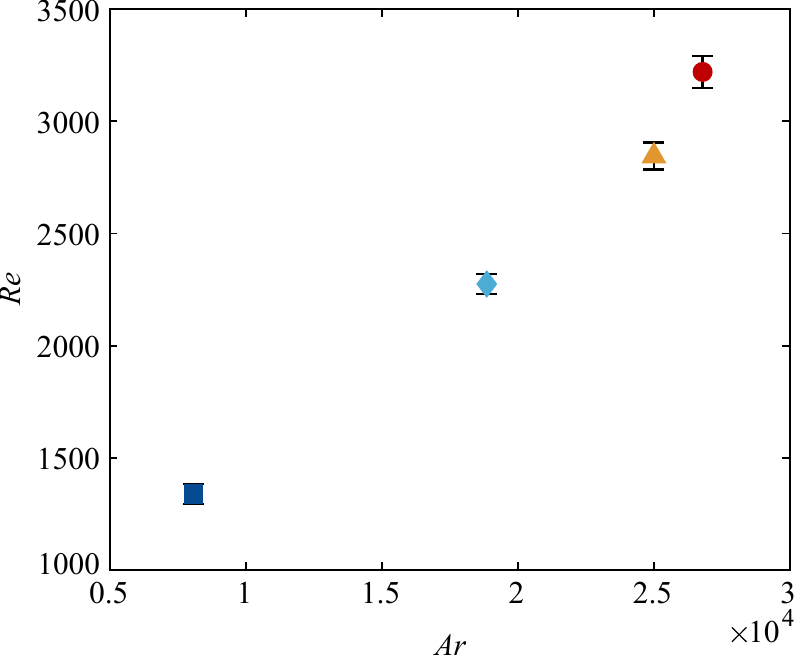}
		\caption{Plot of $Re$ against $Ar$. A linear relation is observed for the first three points and a kink seen for the last, which suggests a transition in settling style. As $Re(Ar)$ is one-to-one in the investigated range, the two are used interchangeably.}
		\label{Fig.Q_ReAr}
	\end{figure}

	In figure \ref{Fig.Q_meanUz} one can see the evolution of the mean descent velocity $\langle V_z \rangle$ with the characteristic lengthscale of the particles $D$. For smaller objects, $\langle V_z \rangle$ decreases as $D$ increases; yet larger particles behave oppositely so a minimum at $D \approx 38\,$mm appears. To examine whether it is related to a change in descent style, the Reynolds number $Re$ is calculated and plotted against the Archimedes number in figure \ref{Fig.Q_ReAr}. The Archimedes number is defined as $Ar = (gD^3 \vert 1-\rho^*\vert)^{\frac{1}{2}}/\nu$, where $g$ is the gravitational acceleration. Previous research has usually observed a linear relation \citep{zhong_experimental_2013, fernandes_zigzag_2005, toupoint_kinematics_2019}, and noted that a change in slope can suggest a transition to another descent style \citep{auguste_falling_2013}. The data in figure \ref{Fig.Q_ReAr} indeed shows a linear relation for the three smallest particles, but there is a modest increase in the slope for the largest particle. This might reflect a physical transition in the particle dynamics, where the upper vertices of the particle with $\theta > 90^\circ$ may interact more with the wake generated by the leading edge. Nonetheless, this feature does not match the local minimum in $\langle V_z \rangle$, whose origin remains unclear.
	
\begin{figure*}
		\centering
		\includegraphics[]{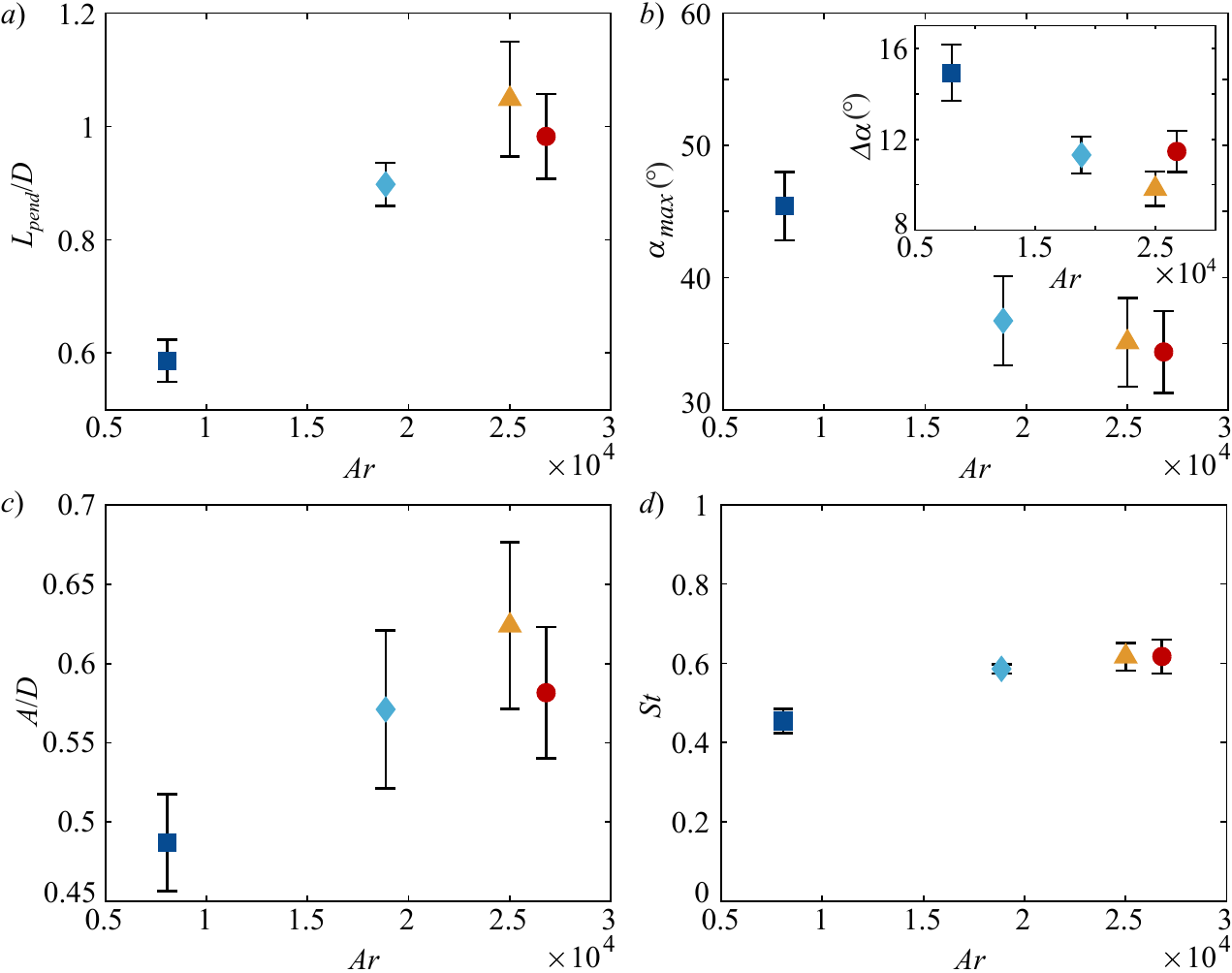}
		\caption{The ({\it a}) dimensionless radius of curvature of the trajectory in a vertical cross-section $L_{pend}/D$, ({\it b}) magnitude of the maximum pitch angle $\alpha_{max}$, (inset) average vertical slip angle over gliding sections $\Delta \alpha$, ({\it c}) dimensionless radial oscillation amplitude $A/D$, and ({\it d}) Strouhal number $St = fD/\langle V_z \rangle$, where $f$ is the radial oscillation frequency, versus the Archimedes number $Ar$.}
		\label{Fig.Q_MinUzExplain}
	\end{figure*}

	Since the descent styles of particles No. 1 through No. 3 appear the same based on the $Ar$--$Re$ plot, we further evaluate the descent velocity behaviour by comparing the radii of curvature of the trajectories $L_{pend}$ in the vertical cross-sections (after applying planar projection and removing the mean descent velocity). We use the subscript `pend' in allusion to the pendulum model that will be introduced later. Similarly, the maximum pitch angles $\alpha_{max}$, the planar oscillation amplitudes $A$ and the dominant radial frequencies $f$ are also evaluated. These are made non-dimensional (except for $\alpha_{max}$) and shown in figure \ref{Fig.Q_MinUzExplain}, where the particles are characterised by their Archimedes number $Ar$. {Note that $A$ differs subtly from $A_g$, which shown in table \ref{Table.DriftResults}, since $A$ includes the turning sections as well.} Both the dimensionless radius of curvature of the particle gliding section $L_{pend}/D$ and amplitude of the oscillations $A/D$ increase with $Ar$. However, for the largest particle, these two magnitudes appear to decrease considerably from the global trend. On the other hand, $\alpha_{max}$ decreases with increasing $Ar$. The Strouhal number, defined as $St = fD/ \langle V_z \rangle$, remains nearly constant across the particles tested, which implies that $f$ is highest for particle No. 1.
	
	Thus, a picture where the smallest particle oscillates rapidly about the vertical axis while descending, and where the larger ones settle more gently emerges. The smallest particle might not be fully gliding, and descends faster with less lift produced. As further evidence, we calculate the average vertical slip angle in the gliding sections {defined as the difference between the pitch angle and the angle of inclination of the velocity vector, i.e.
		\begin{equation}
		\Delta \alpha = \biggl\langle \tan^{-1}\biggl[\frac{V_z}{(V_x^2 + V_y^2)^{1/2}}\biggr] - \vert\alpha\vert \biggr\rangle.
		\end{equation}
		This is plotted in the inset of figure \ref{Fig.Q_MinUzExplain}\,({\it b\/}). The figure shows that $\Delta \alpha$ decreases slowly as the particle diagonal length $D$ increases for particle No. 1 to 3, therefore proving its pitch attitude is more closely aligned with the velocity vector.}
	
	Indeed, such a difference in falling behaviour can explain the initial reduction of the mean descent velocity at small $Ar$. When the particle's curved surface area increases, more lift is generated and the gliding motion is enhanced, leading to a reduction in $\langle V_z \rangle$. However, this argument alone cannot explain the minimum in $\langle V_z \rangle$.
	
\begin{figure}
		\centering
		\includegraphics[]{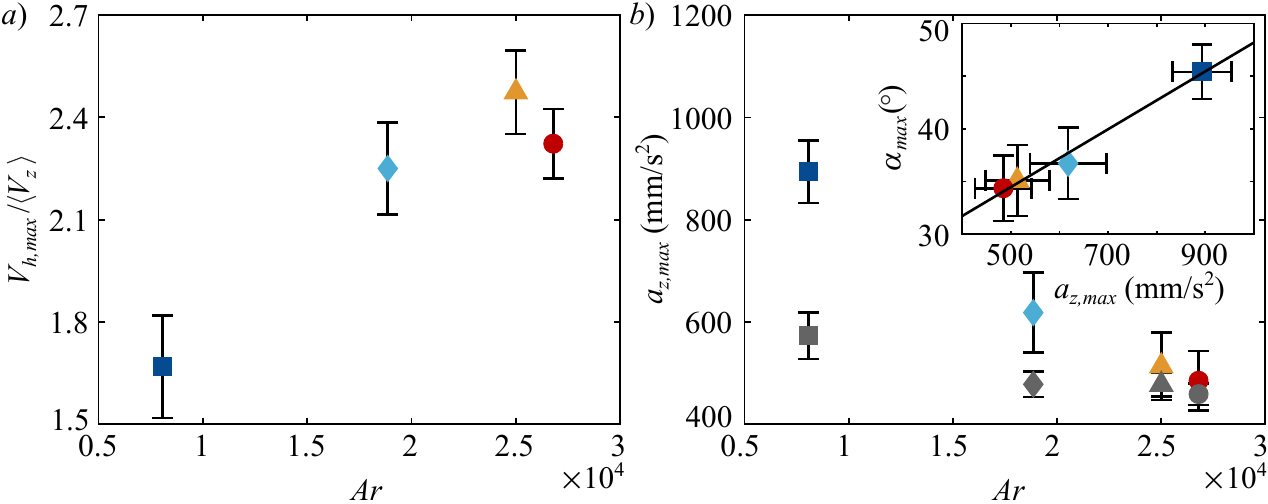}
		\caption{({\it a}) Plot of the non-dimensional maximum horizontal speed in a swing $V_{h,max}/\langle V_z \rangle$. ({\it b}) Behaviour of the maximum vertical acceleration $a_{z,max}$. {The grey symbols represent the descending pendulum model introduced in \S \ref{Sec.PendModel}.} Inset shows $\alpha_{max}$ versus $a_{z,max}$ together with the best-fit line: {$\alpha_{max} = (0.03\pm0.01)a_{z,max} + (20.8\pm6.8)$.}}
		\label{Fig.Q_UhmaxAlpha}
	\end{figure}

	To understand why the descents become faster at larger $Ar$, we measure the maximum horizontal speed in each swing $V_{h,max}$. As figure \ref{Fig.Q_UhmaxAlpha}\,({\it a\/}) illustrates, {$V_{h,max}/\langle V_z \rangle$ generally grows with $Ar$. This increase in $V_{h,max}$ leads to a larger $\langle V_z \rangle$ because the particle pitches down at the beginning of each swing, so the horizontal and vertical speeds are coupled to each other. Therefore, the minimum in the descent velocity manifests through} a delicate balance between lift enhancement and a reduction of the particle's horizontal speed during the glide.
	
	Settling behaviour at large $Ar$ (or equivalently $\theta$) is more complex. While $\langle V_z \rangle$ increases even for the largest object, the behaviour of particle No. 4 is different from the other ones. Figure \ref{Fig.Q_MinUzExplain}\,({\it a,\,c\/}) demonstrate that $L_{pend}/D$ and $A/D$ are reduced as compared with the linear extrapolation from the previous three. The trend in $A/D$ could be related to the increase in the slope of $Re(Ar)$. If energy is conserved, a smaller $A/D$ implies more potential energy is converted to vertical velocity instead of horizontal velocity. Since $Re$ is based on $\langle V_z \rangle$, the $Re(Ar)$ relation becomes steeper as argued in \cite{auguste_falling_2013}. We hypothesise that the differences observed are due to stronger interactions between the leading edge vortex and the upper vertices of the particle. Further work is required to understand this behaviour. 
	
	Since $\alpha_{max}$ indirectly determines the position of the slowest descent, linking it to a more experimentally accessible quantity might be useful. {As discussed, the larger particles oscillate with a smaller $\alpha_{max}$ and descend more smoothly. This is also reflected by the maximum vertical acceleration $a_{z,max}$ displayed in figure \ref{Fig.Q_UhmaxAlpha}\,({\it b\/}). The inset shows} $\alpha_{max}$ is linearly related to $a_{z,max}$. This is somewhat expected for the gliding particles since a larger initial pitch angle would mean a steeper descent near the extrema. However, it is worth noting that the same slope extends to even the smallest particle which settles without generating significant lift based on our interpretation.
	
	\subsection{Modelling the settling behaviour}	\label{Sec.PendModel}
\begin{figure}
		\centering
		\includegraphics[]{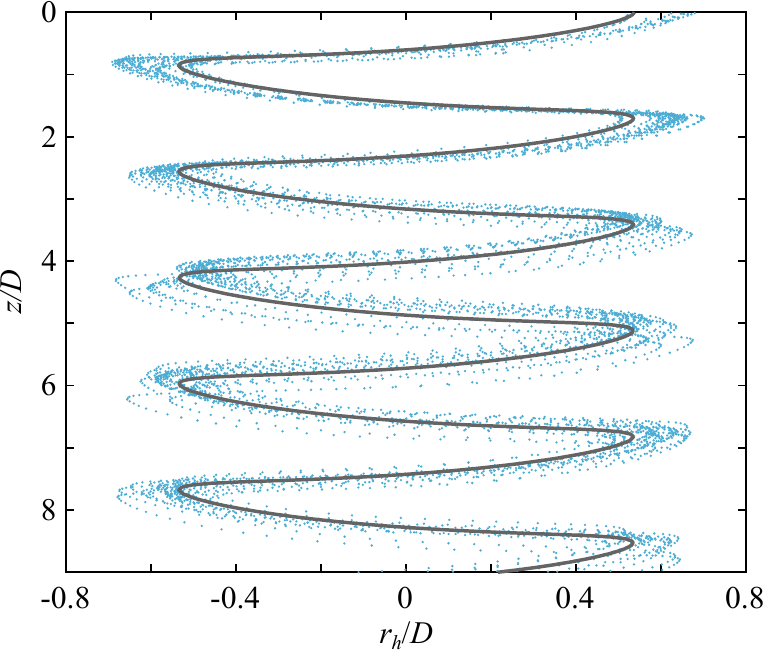}
		\caption{Model pendulum trajectory {(grey line)} overlaid on the experimental data of particle No. 2.}
		\label{Fig.Q_PendTraj}
	\end{figure}

	As the particles oscillate periodically while settling, pendulums {whose pivots descend at constant speeds are chosen to model their motions, as also proposed for freely falling disks \citep{esteban_dynamics_2019}.} Motivated by the fact that the amplitude of the motion does not vary in time as the particle settles (see figure \ref{Fig.Q_Th73OneTraj}\,{\it b\/}), an idealized pendulum model is constructed assuming that the system is non-dissipative. Thus, their equation of motion ignoring the constant vertical descent velocity reads
	\begin{equation}	\label{Eq.Pendulum}
	\frac{d^2\phi}{dt^2} = -\frac{g(\rho^*-1)C}{L}\sin\phi
	\end{equation}
	where $\rho^* > 1$. Here, $\phi$ is the angular displacement from the vertical, $L$ the (virtual) pendulum length --- i.e. the length from the swinging particle to the virtual origin falling vertically with the particle --- and $C$ a constant to account for all accelerations apart from gravity. By definition, $L$ and the initial angular position correspond to the measured quantities $L_{pend}$ and $\alpha_{max}$ respectively. This leaves only $C$ as a fitting parameter, whose value is found by matching the oscillation frequencies to experiments. {Although previous studies \citep{tanabe_behavior_1994,belmonte_flutter_1998} have used pendulums to describe the dynamics of settling particles, they focus on the quasi-two-dimensional scenario involving flat plates as opposed to our fully three-dimensional case with curved particles.
		
		Figure \ref{Fig.Q_PendTraj} shows the pendulum trajectory overlaid on the experimental data of particle No. 2. Although not shown, similar plots are obtained for all four particles. In view of the reasonably good agreement between the experimental data and the model proposed, `planar zig-zag' descents can be viewed as simple harmonic motions superposed on uniform descents, though higher-order quantities such as $a_{z,max}$ are not accurately captured by the model as indicated by the grey symbols in figure \ref{Fig.Q_UhmaxAlpha}\,({\it b\/}).}
	
\begin{figure}
		\centering
		\includegraphics[]{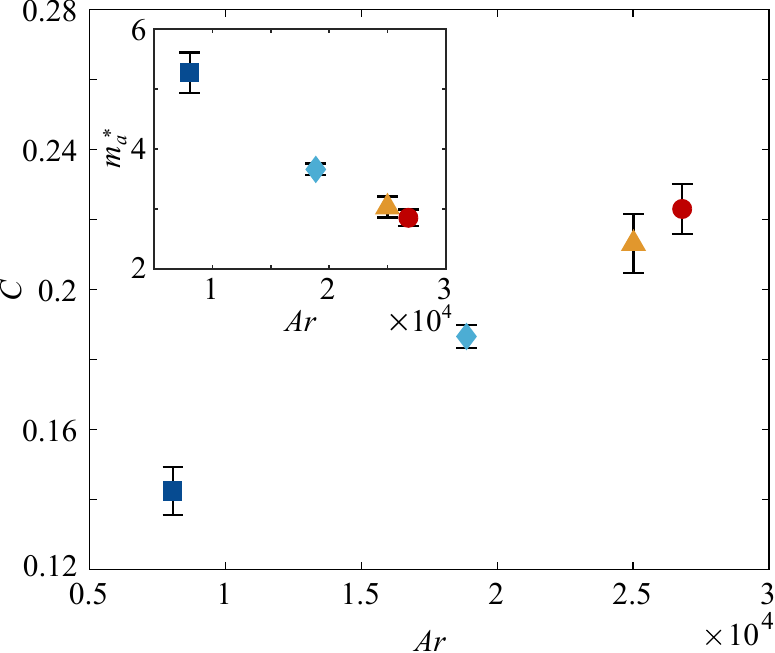}
		\caption{Fitting parameter $C$ as a function of $Ar$. In the context of underwater pendulums, $C$ is related to the added-mass coefficient $m_a^*$. Inset shows $m_a^*$ versus $Ar$.}
		\label{Fig.Q_PendParameters}
	\end{figure}

	To examine whether the fitted parameter $C$ can be determined without frequency matching, it is plotted against $Ar$ in figure \ref{Fig.Q_PendParameters}. Interestingly, a nearly linear relation exists, meaning this model allows one to predict the particle velocity fluctuations simply by computing $Ar$ without any \textit{a priori} knowledge. In the context of undamped underwater pendulums, $C = (\rho^* + m_a^*)^{-1}$, where $m_a^*$ is the added-mass coefficient characterising the energy spent accelerating the surrounding fluid. This can be obtained by comparing (\ref{Eq.Pendulum}) with the equation of motion of underwater pendulums as in \cite{mathai_dynamics_2019},
	\begin{equation}
	\frac{d^2\phi}{dt^2} = -\frac{g(\rho^*-1)}{L(\rho^*+m_a^*)}\sin\phi.
	\end{equation}
	The inset of figure \ref{Fig.Q_PendParameters} shows that $m_a^*$ decreases with increasing $Ar$, suggesting that enhanced gliding means less effort is required to move the neighbouring liquid. The magnitude of this parameter is much larger than in objects like cylinders since particle volume and $\rho_f$ are used to non-dimensionalize the added mass.
	
	To better understand the behaviour of bottle fragments in industrial facilities, the same objects are dropped in the water tank with background turbulence. In the following section, the flow characteristics are presented and the dynamics of the particles discussed.
	
	\section{Settling in turbulent flow}	\label{Sec.Turb}
	The experiments are conducted in a random jet array facility (figure \ref{Fig.Setup}\,{\it e\/}), where turbulence is generated by the continuous action of submerged water pumps as in \cite{esteban_laboratory_2019}. However, the addition of a pulse-width-modulation system allows us to control turbulence intensity while the facility is in operation. The characteristics of the turbulence generated are in table \ref{Table.TurbStat} and further details can be found in Appendix \ref{Sec.Turb_Gen}. Turbulence is produced so that all the particles' characteristic {length scales are smaller than the horizontal integral length scale $L_x$. As mentioned in \S \ref{Sec.QuiescentMethods}, these particles have sizes comparable to the ones processed in actual recycling facilities.} The experimental procedure to release particles in this section is analogous to the one previously presented. However, as turbulent flow quantities can only be predicted statistically, the {number of repeated experiments per particle is increased to at least 49}. The minimum waiting time between releases is reduced to 3 minutes since the background turbulence washes residual flows away rapidly. Nonetheless, to ensure statistical stationarity, the pumps are switched on for no less than 10 minutes before the first drop. {We position the lower camera further back which resulted in a resolution of $\approx 0.35$\,mm/px and a magnification of $\approx 1/60$.} We also monitor the water temperature for accurate estimation of the dimensionless parameters.
	
	\begin{table}
		\centering
		\begin{tabular}{l@{\hskip 1cm}l}
			Turbulence statistics & \multicolumn{1}{c}{Values}\\
			$\overline{u_x}/\overline{u_z}$ & 1.34\\
			$u'_{rms} = (\overline{u_x}+2\overline{u_z})/3$ & $15.9\,$mm/s\\
			$MFF$ & 0.48\\
			$HD$ & 0.07\\
			$L_{turb}$ & $45.0\,$mm\\
			\hspace{5mm}$(L_x,L_z)$ & \hspace{5mm}$(65.3,34.8)\,$mm\\
			\hspace{8mm}$(L_{xx},L_{zx})$ & \hspace{8mm}$(93.9,45.4)\,$mm\\
			\hspace{8mm}$(L_{zz},L_{xz})$ & \hspace{8mm}$(44.3,27.4)\,$mm\\
			($\lambda_f$,$\lambda_g$) & $(6.8,6.5)\,$mm\\
			\hspace{5mm}$(\lambda_{xx},\lambda_{zx})$ & \hspace{5mm}$(8.1,7.8)\,$mm\\
			\hspace{5mm}$(\lambda_{zz},\lambda_{xz})$ & \hspace{5mm}$(6.1,5.8)\,$mm\\
			$Re_\lambda$ & 98\\
			\hspace{5mm}$(Re_{\lambda,x},Re_{\lambda,y})$ & \hspace{5mm}$(139,77)$\\
		\end{tabular}
		\caption{Statistics of the background turbulence such as the root-mean-square velocity fluctuations, the mean flow factor ($MFF$), homogeneity deviation ($HD$), integral lengthscales and Taylor microscales. $\lambda_f$ and $\lambda_g$ denote the longitudinal and transverse Taylor microscales, respectively. The reader is referred to the appendix for the full definitions. The values in brackets correspond to the respective quantities in the column on the left.}
		\label{Table.TurbStat}
	\end{table}
	
	Data analysis is very similar to the cases in quiescent fluid, with the main differences being the identification of the transients and that the trajectories are no longer detrended to account for horizontal drifts. The presence of background turbulence means any transient effects are confined to an even smaller section of the trajectory. Despite this, for each descent in turbulence, we still remove the mean length of the transients for the corresponding quiescent experiments from the trajectory.
	
	\subsection{Results and discussion}
	Several particle descents in turbulence are plotted in figure \ref{Fig.TurbTraj} (see the supplementary videos). The `planar zig-zag' mode found in quiescent fluid is still present, with the dominant oscillation frequency over each trajectory nearly unchanged in all particles tested. However, their motions are diversified by flow fluctuations and therefore trajectories are no longer repeatable. Still, four types of special events are identified across all the particles investigated: 1) `slow descents', where the quiescent settling style remains but vertical velocity is attenuated (figure \ref{Fig.TurbTraj}\,{\it a\/}); 2) `rapid rotation', where the direction of the oscillations changes rapidly at the end of a swing (figure \ref{Fig.TurbTraj}\,{\it b\/}); 3) `vertical descents', where the planar motion diminishes and the particle essentially falls straight down (figure \ref{Fig.TurbTraj}\,{\it c\/}); and 4) `long gliding motions', where the gliding section in the `planar zig-zag' motion is especially long {($\approx 4.8D$ in the illustrated case)} and is sometimes preceded by a large $\alpha$ (figure \ref{Fig.TurbTraj}\,{\it d\/}). Apart from vertical descents, which we do not observe for particle No. 1, these events occur for all the particles. Multiple types of the motions listed may occur in a single descent. Remarkably, the particles never flip over, possibly due to their dihedral configuration.
	
	\begin{figure}
		\centering
		\includegraphics[]{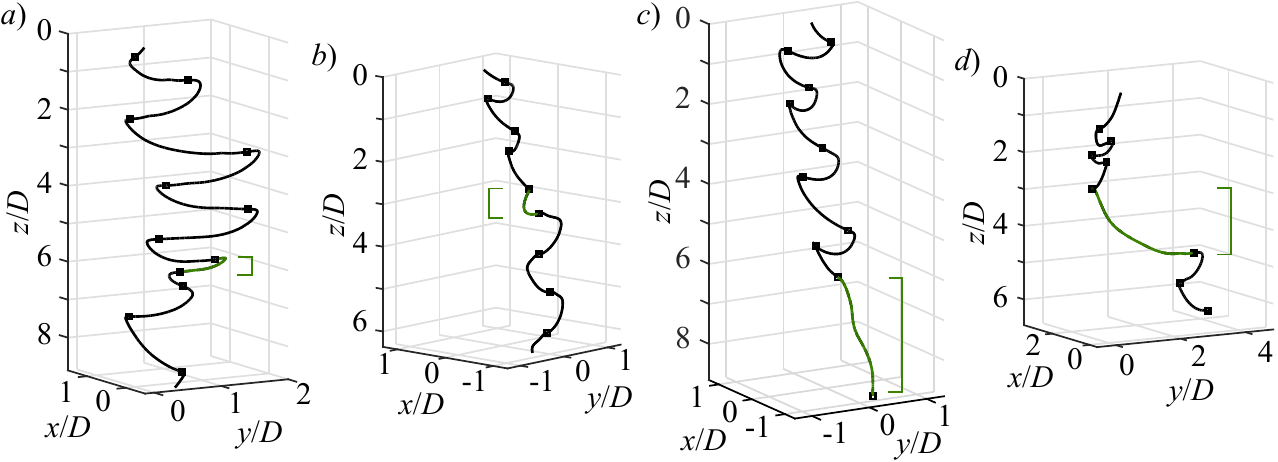}
		\caption{Trajectories of particles in turbulent flow. Four special types of motions are observed though the underlying zig-zag mode seen in quiescent fluid remains: ({\it a}) `slow descent', ({\it b}) `rapid rotation', ({\it c}) `vertical descent'  and ({\it d}) `long gliding motion'. The positions of the events within the trajectories are marked by square brackets on the side. Square markers denote locations corresponding to local minima of $V_z$.}
		\label{Fig.TurbTraj}
	\end{figure}
	
	Slow descents probably occur when the object encounters strong incident flows that enhance lift. As the smallest particle does not generate sufficient lift to fully glide in quiescent fluid, it indeed rarely exhibits this behaviour. Rapid rotations can emerge when the solid enters a region of horizontal shear, causing it to rotate and sometimes roll slightly. This kind of motion becomes more likely the smaller the $I_{roll}$ or the larger the distance between the centre-of-gravity and the centre-of-pressure (i.e. a longer moment arm). Heuristically, assuming the centre-of-pressure coincides with the centre of the solid's circle of curvature when viewed at the front (figure \ref{Fig.Setup}\,{\it b\/}), the smallest particle has the longest moment arm. Either way, the smallest object should be the most sensitive to such shear. Long gliding motions appear possibly as the local background flow has a significant component along the particle's direction of motion, pushing it along. Finally, we hypothesise that vertical descents happen when the object encounters a downdraught.
	
	Slow descents and long gliding motions have also been found for disks falling under background turbulence in \cite{esteban_disks_2020}. However, we noticed key distinctions in the settling characteristics between these two geometries. First, rapid rotations have not yet been observed for disks. Second, fast descents of disks differ from vertical descents of the particles tested here. This type of motion for disks is always preceded by an especially large $\alpha$, so the disks are aligned with the direction of motion. However, this is not necessarily the case for the bottle-fragment-like particles.
	
\begin{figure}
		\centering
		\includegraphics[]{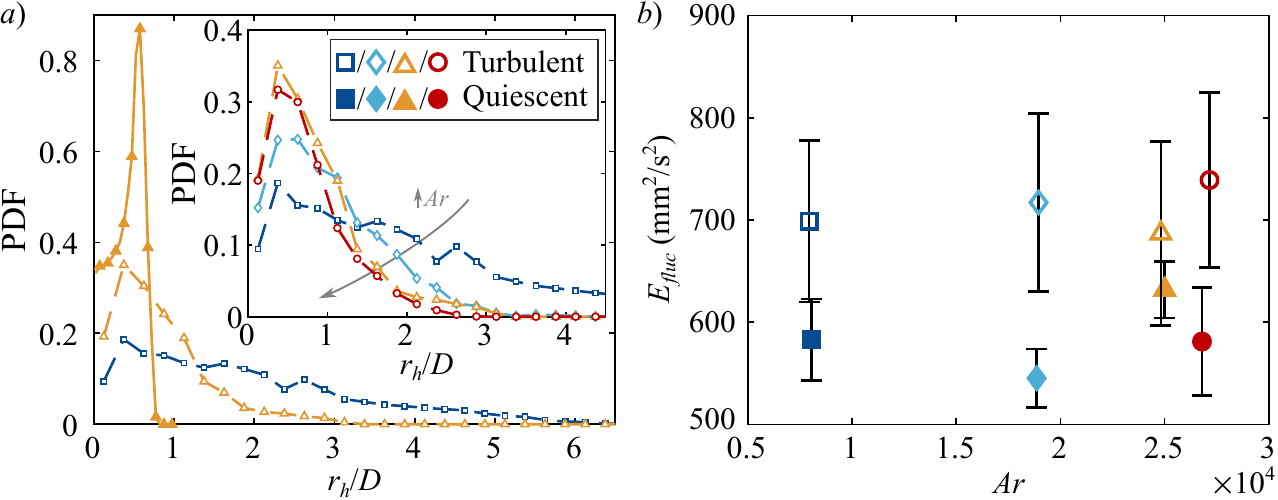}
		\caption{({\it a}) Probability density functions (PDF) of $r_h/D$ along the descent. The solid line and solid data points are for quiescent fluid while the dotted one and hollow data points are for turbulent settling. PDF of particles No. 1 {(turbulent case only)} and No. 3 are displayed. The inset shows the same quantity but for all the objects dropped. The symbols follow those introduced in table \ref{Table.ParticleProp}. ({\it b}) Vertical fluctuation kinetic energy per unit mass $E_{fluc}$ of the various particles.}
		\label{Fig.PDFrKEfluc}
	\end{figure}

	To assess the effect of turbulence on all the descents collectively, the height-integrated radial probability density functions (PDF) and the specific kinetic energy fluctuations of $V_z$ (i.e. half of the variance of fluctuations of $V_z$), $E_{fluc}$, are shown in figure \ref{Fig.PDFrKEfluc}. To accurately capture the radial displacement $r_h$, non-transient parts of the trajectories are centred so the origin coincides with the mean position of the first swing.
	
	The diversification of the settling dynamics by background turbulence is also evident here. For the horizontal motion, focusing first on figure \ref{Fig.PDFrKEfluc}\,({\it a\/}), the radial PDF in quiescent and turbulent flows of particle No. 3 reveal that the most likely radial position remains unchanged. This confirms that the quiescent zig-zag motion is still significant at the current turbulence level. Yet, the PDF is now much broader, with particle dispersion reaching multiple $D$ instead of only $r_h/D \approx 1$. The inset in figure \ref{Fig.PDFrKEfluc}\,({\it a\/}) shows how the radial dispersion of the particles in turbulence reduces as $Ar$ increases. However, the vertical component of the velocity fluctuations are modified differently. These are shown in figure \ref{Fig.PDFrKEfluc}\,({\it b\/}), and demonstrate a strong increase in velocity fluctuations about $\langle V_z \rangle$. {Hence, the motion is destabilised to a similar extent over most $Ar$ tested. This difference may be attributed} to gravity, which has been used by \cite{byron_slip_2019} to explain an identical trend for slip velocities of nearly neutrally buoyant cylinders in turbulence.
	
\begin{figure}
		\centering
		\includegraphics[]{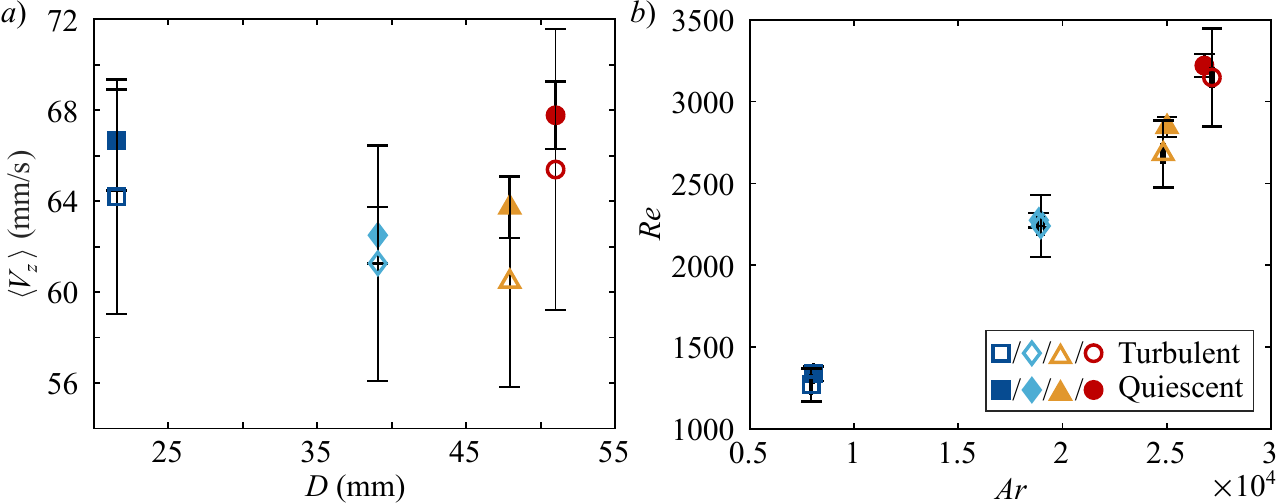}
		\caption{({\it a}) Mean descent velocities $\langle V_z \rangle$ of the various particles in turbulence and quiescent fluid. ({\it b}) Dimensionless version of the descent velocity plot, $Re(Ar)$.}
		\label{Fig.TurbUz}
	\end{figure}

	The effect of turbulence on the mean descent velocity $\langle V_z \rangle$ has long been an area of great interest. Figure \ref{Fig.TurbUz}\,({\it a\/}) plots $\langle V_z \rangle$ against the particle characteristic lengthscale, showing $\langle V_z \rangle$ reduces compared with the quiescent case, although the data lies within the statistical deviation of the turbulent one. We note this result is congruent with the slip velocity of nearly neutrally buoyant cylinders \citep{byron_slip_2019}, and opposite to $\langle V_z \rangle$ of inertial disks falling in background turbulence \citep{esteban_disks_2020}. {As mentioned in \S \ref{Sec.Intro}, \cite{good_settling_2014} found that settling is hindered by turbulence when the characteristic gravitational velocity is greater than the typical flow velocity fluctuations $u'_{rms}$. To compare this with our results, we formed an analogous quantity by replacing the characteristic gravitational velocity with the mean descent velocity in quiescent fluid $\langle V_q \rangle$. $\langle V_q \rangle/u'_{rms}$ is found to lie in between $3.92$ and $4.27$. Hence, our results are in agreement with the prediction by \cite{good_settling_2014} which suggests that the mean descent velocity would be reduced when $\langle V_q \rangle/u'_{rms} > 1$. We recognise $\langle V_q \rangle/u'_{rms}$ does not reflect the multiscale nature of particle--turbulence interactions, and it may be more insightful to employ a scale-dependent quantity instead. However, theoretically deriving such a quantity for our particle geometry is highly non-trivial and is beyond the scope of this study.}
	
	{To further investigate the cause of the hindered settling,} the relation between $Re$ and $Ar$ is shown in figure \ref{Fig.TurbUz}\,({\it b\/}). The general trend observed is the same as in quiescent fluid --- with an approximately linear relation for the three smallest particles and an increase in slope for the last one --- and an identical interpretation is employed. As considering quantities averaged over entire trajectories do not seem to help explain the change in $\langle V_z \rangle$, particle motions are examined over trajectory sections. \cite{esteban_disks_2020} studied the correlation between $\langle V_z \rangle$ and the dominant frequency of each trajectory. Instead of following this approach, where the existence of a single `weak' event might be hidden by the presence of more severe ones, we propose an alternative method to capture the effect of all the events, the average descent velocities $V_{event}$ and the characteristic frequencies $f_{event}$ conditioned on each type of event. However, this leads to a practical question on the definition of an `event'.	
	
\begin{figure}
		\centering
		\includegraphics[]{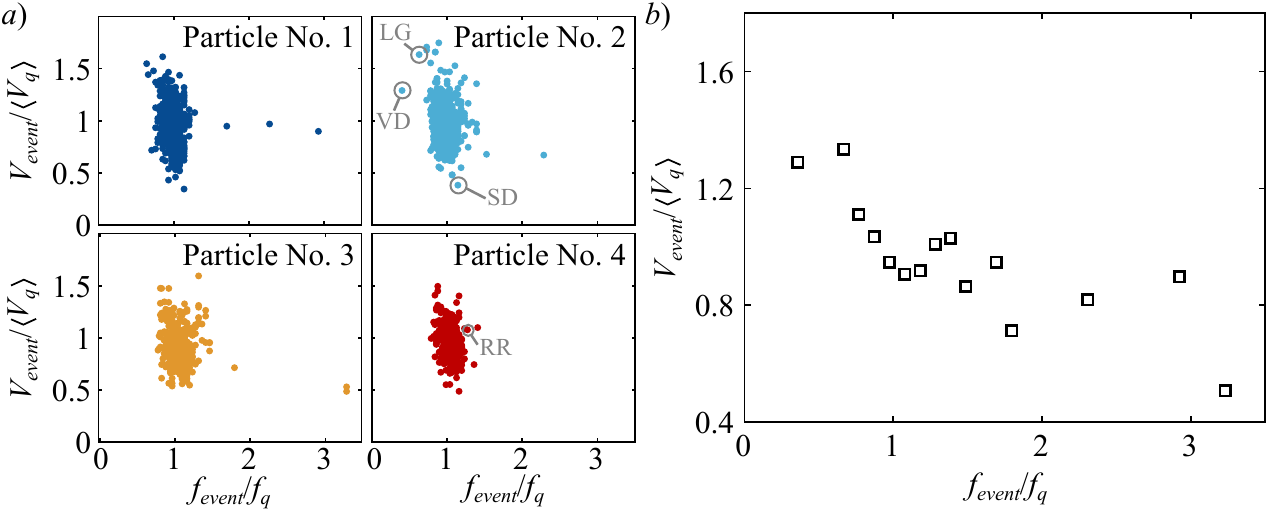}
		\caption{{({\it a})} Scatter plots showing the relationship between the average descent velocities $V_{event}$ and frequencies $f_{event}$ per event. To better visualise the effect of background turbulence, these quantities are normalised by the mean descent velocity $\langle V_q \rangle$ and the dominant vertical oscillation frequency $f_q$ in quiescent fluid. The points corresponding to the special events shown in figure \ref{Fig.TurbTraj} --- slow descent (SD), rapid rotation (RR), long gliding motion (LG) and {vertical descent (VD)} --- are annotated. {({\it b}) The average $V_{event}$ against $f_{event}$ of all the events in ({\it a}).}}
		\label{Fig.EventFreq}
	\end{figure}

	Classifying events using the instantaneous vertical velocity provides reasonable results. The positions corresponding to local minima of $V_z$ (squares in figure \ref{Fig.TurbTraj}) also match those of the radial extrema reasonably well, and these are used to separate events. Each event then essentially corresponds to a half-swing, with $f_{event}$ being the inverse of its duration. Figure {\ref{Fig.EventFreq}\,({\it a})} shows the mean descent velocity of each event $V_{event}$ versus $f_{event}$, both normalised by the corresponding mean values in quiescent liquid.
	
	In general, events with small frequencies $f_{event}$ can increase the descent velocity $\langle V_z \rangle$, while those with large $f_{event}$ have the opposite effect. {This is quantitatively illustrated by figure \ref{Fig.EventFreq}\,({\it b}) where the mean event velocity $\langle V_{event} \rangle$ is plotted against $f_{event}$.} The same {was} also found for disks in \cite{esteban_disks_2020}, although the trend here is less prominent due to the moderate particle inertia. Also, particle {No. 2 exhibits a wider variety of events compared to particle No. 1} as reflected by the scatter in the data, in agreement with the initial observation that certain types of motions are less frequent for smaller particles. Contrary to the variation in the horizontal displacement (see figure \ref{Fig.PDFrKEfluc} {\it a\/}), turbulence introduces more extreme events for the larger particles.
	
\begin{figure}
		\centering
		\includegraphics[]{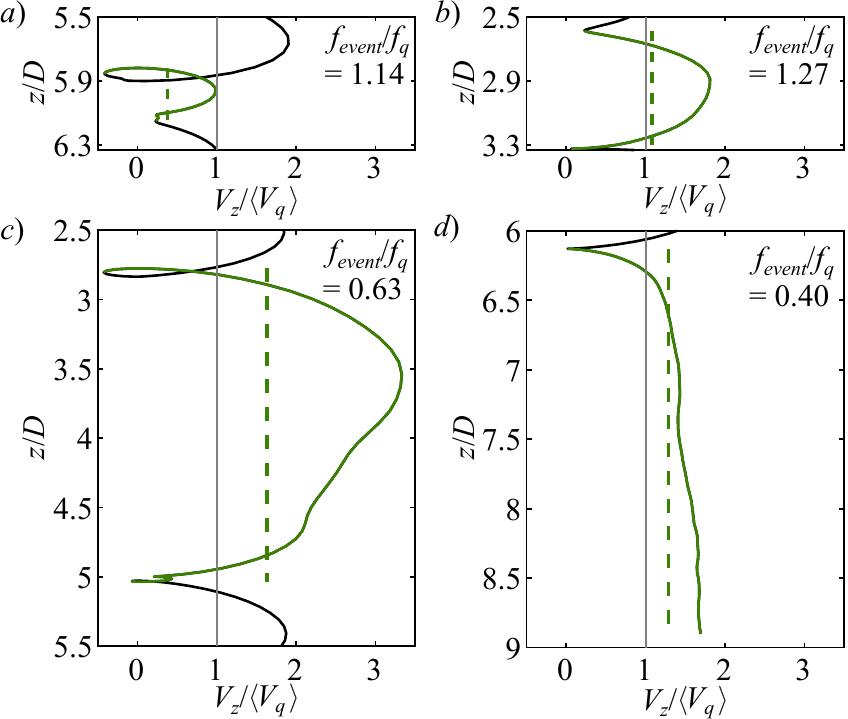}
		\caption{The instantaneous descent velocities $V_z$ (solid lines) and the mean values $V_{event}$ (dotted lines) of the four types of events identified: ({\it a}) slow descent, ({\it b}) rapid rotation, ({\it c}) long gliding motion and ({\it d}) vertical descent. The green lines show the locations of the events.}
		\label{Fig.TurbEventUz}
	\end{figure}

	To correlate the four types of events with the modulation in frequency, figure \ref{Fig.TurbEventUz} displays the variation of $V_z$ over their durations, the corresponding $V_{event}$ and $f_{event}$. Each type of descent behaviour modifies $\langle V_z \rangle$ differently: `slow descents' have $V_{event} \approx 0.5\langle V_z \rangle$ (figure \ref{Fig.TurbEventUz}\,{\it a\/}); rapid rotations have no significant effects on $\langle V_z \rangle$ (figure \ref{Fig.TurbEventUz}\,{\it b\/}), meaning the rotation is not coupled to the vertical motion; long gliding motions (figure \ref{Fig.TurbEventUz}\,{\it c\/})  could considerably enhance settling, regardless of the initial pitch angle; vertical descents (figure \ref{Fig.TurbEventUz}\,{\it d\/}) increase $\langle V_z \rangle$. The behaviour of vertical descents is as expected since the distance travelled is shorter compared to zig-zag, and downdraughts force the particle down{. Although} the limited depth of our tank means the vertical descent in figure \ref{Fig.TurbEventUz}\,({\it d\/}) is incomplete, we are confident that the complete event still increases $\langle V_z \rangle$ for the reasons above. In summary, as long gliding motions and vertical descents have small $f_{event}/f_q$, they correspond to points with small $f_{event}$ and large $V_{event}$ in figure \ref{Fig.EventFreq}\,{({\it a})}.
	
\begin{figure}
		\centering
		\includegraphics[]{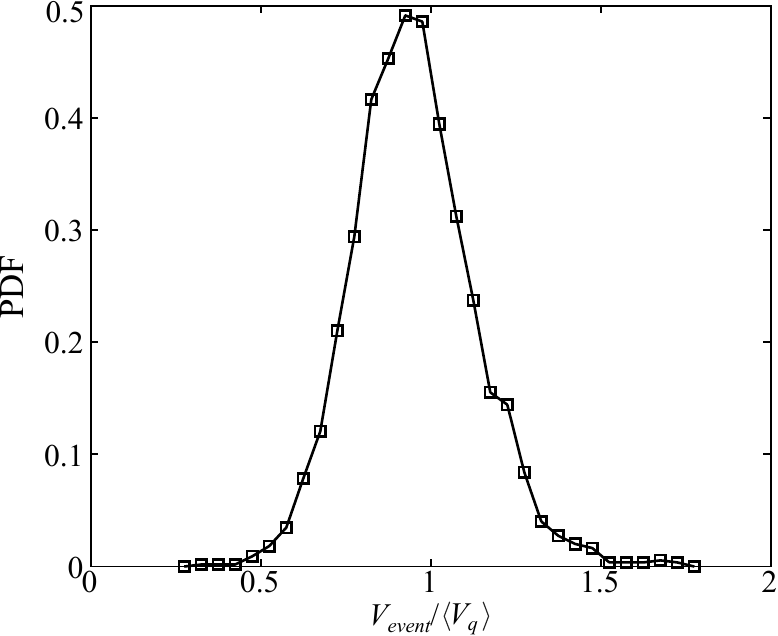}
		\caption{PDF of $V_{event}$ relative to $\langle V_q \rangle$. To avoid overcounting slow events, the successive ones with $V_{event} < 0.4 \langle V_q \rangle$ have been merged.}
		\label{Fig.PDFUevent}
	\end{figure}

	So far, it has been shown that low-frequency events such as long gliding motions and vertical descents could enhance settling. However, $\langle V_z \rangle$ is smaller than the quiescent value on the whole. This result is captured when plotting the PDF of $V_{event}$ (figure \ref{Fig.PDFUevent}). Before proceeding, note that the definition of events used may over-count the slow ones. This is mitigated by combining successive events with $V_{event} < 0.4 \langle V_q \rangle$. Though the threshold is somewhat arbitrary, it does not affect the following discussion. The reduction in $\langle V_z \rangle$ is manifested as a slight leftward shift of the entire PDF.
	
	{Among the four types of events identified, only slow descents reduce the settling speed. However, we recognise that the events described are the most readily detected ones and do not constitute an exhaustive list. Particle settling in turbulence is a highly complex and multiscale phenomenon \citep{tom_multiscale_2019} that exhibits a number of more subtle unclassified events. We therefore believe the attenuation in settling may be caused by the less discernible events.} As the falling particle resembles a swept back wing in the direction of motion and larger particles glide more in quiescent fluid, it is possible that under most conditions, the turbulence provides slightly more lift without considerably changing the basic zig-zag motion.
	
	Admittedly, such a result is unexpected. Since the particle sizes are of the same order as the integral lengthscale $L_{turb}$, we anticipated the solid to exhibit downward sweeping motions triggered by interactions with large vortices. However, the object's inherent stability likely suppresses these motions.
	
	\section{Concluding remarks} \label{Conclusion}
	Motivated by the numerous applications of particle settling, such as differentiating plastic from glass in hydrodynamic separators, 3D-printed rigid thin curved solids resembling bottle fragments were dropped in a water tank in quiescent fluid and in homogeneous anisotropic turbulence.
	
	In quiescent liquid, the particles underwent planar zig-zag descent and their trajectories were divided into gliding and turning sections. While one might expect the average vertical velocity $\langle V_z \rangle$ to vary monotonically with particle size, a minimum was found at $D \approx 38\,$mm ($Ar \approx 1.8\times10^4$). Closer examination of the settling behaviour showed that the horizontal oscillation amplitude $A$ and radius of curvature $L_{pend}$ normalised by particle size were generally enhanced for larger particles. On the contrary, the oscillation frequency $f$ and the maximum pitch angle $\alpha_{max}$, which was directly proportional to $a_{z,max}$, decreased monotonically. These suggested enhanced lift generation as the particle size grew, which was supported by a {closer} alignment between $\alpha$ and the direction of motion. This led to the initial reduction in $\langle V_z \rangle$. The subsequent settling enhancement for the larger objects was due to more rapid horizontal motion at mid-swing locations coupled with their initial pitch down attitude at the beginning of each swing. All the trajectories observed could be modelled reasonably well by undamped underwater pendulums descending at a constant velocity.
	
	The zig-zag motion was also observed for settling in turbulence, but fluctuations in the flow modified it so the radial dispersion increased considerably. Notably, the particles never flipped over although their sizes were comparable to $L_{turb}$. In agreement with \cite{good_settling_2014}, $\langle V_z \rangle$ was slightly lower than in quiescent fluid for $\langle V_q \rangle/u'_{rms}>1$. Four special types of events comprising slow descents, vertical descents, long gliding motions which were sometimes preceded by large pitch angles, and rapid rotations, were identified. Also, each type of motion was related to the particle kinematics and to the descent velocity. In general, vertical descents and long gliding sections sped up settling. By dividing each trajectory into a collection of events, those with a low frequency were found to be capable of enhancing the descent, while the opposite occurred for high-frequency events. Nevertheless, the PDF of $V_{event}$ was unimodal and the reduction of $\langle V_z \rangle$ was reflected by a leftward {shift}. This may suggest the background flow slightly modulated each event by enhancing lift production, so the change in $\langle V_z \rangle$ could not be simply connected to the special events. The above also underlines the difficulty of studying descent behaviour with background turbulence.
	
	Future research may therefore focus on wake visualisation of these particles in both turbulence and quiescent fluid. As transitions in settling behaviour are usually correlated to a change in wake structure (see e.g. \cite{ern_wake-induced_2012, lee_experimental_2013, auguste_falling_2013, esteban_three_2019, toupoint_kinematics_2019}) and $\alpha_{max} \propto a_{z,max}$ found here indirectly supports this argument, observing the wake {may reveal other types of events and the effects of anisotropic geometries. This may} further explain the change of $\langle V_z \rangle$ in turbulence and the lift enhancement as the particle size increased in quiescent fluid. Moreover, it may uncover why certain trends reversed for particle No. 4, where $\theta > 90^\circ$. {Theoretical development may concentrate on finding a suitable scale-dependent metric for anisotropic particles to distinguish between enhanced and hindered settling in turbulence. Finally, additional development of the pendulum model is desirable.} An emphasis should be placed on interpreting $C$ as it may complement the current experimental observations and improve the predictive power of the model.
	
	\section*{Acknowledgements}
	We thank Jelle Will for fruitful discussions and Dominik Krug for drawing our attention to literature which modelled settling behaviour with pendulums. T.T.K.C. also thanks the Internship Office at the University of Twente and the Faculty Office of the Faculty of Engineering and Physical Sciences at the University of Southampton. He is partially funded by the University of Twente Scholarhip and the Erasmus\textsuperscript{+} Traineeship Scholarship. S.G.H. acknowledges MCEC for financial support.
	
	\section*{Declaration of interests}
	The authors report no conflict of interest.
	
	\section*{Supplementary materials}
	Five supplementary videos showing the descents depicted in figure \ref{Fig.Q_Th73OneTraj} (`Movie 1.mp4') and figure \ref{Fig.TurbTraj} (`Movie 2.mp4', `Movie 3.mp4', `Movie 4.mp4' and 'Movie 5.mp4' corresponding to figure \ref{Fig.TurbTraj}\,\textit{a}--\textit{d} respectively) complement this paper.
	
	\appendix
	\section{Turbulence generation and characteristics}
	\label{Sec.Turb_Gen}
	As explained in \S \ref{Sec.Turb}, the experiments are conducted in a random jet array facility (figure \ref{Fig.Setup}\,{\it e\/}), where turbulence is generated by the continuous action of submerged water pumps. These pumps, arranged in two 8 $\times$ 6 arrays with vertical and horizontal mesh lengths of $10\,$cm on either side of the tank, fire independently according to the `Sunbathing Algorithm' to generate statistically stationary homogeneous anisotropic turbulence with negligible mean flow \citep{variano_random-jet-stirred_2008, esteban_laboratory_2019}. The duration of the `on' and `off' signals are randomly selected from two separate Gaussian distributions with their mean values and standard deviations denoted by $\mu_{on/off}$ and $\sigma_{on/off}$ respectively. In this case, $\mu_{on} \pm \sigma_{on} = (3 \pm 1)\,$s and $\mu_{off} \pm \sigma_{off} = (21 \pm 7)\,$s. When the pumps are active, water is drawn radially at their bases and expelled horizontally out of their cylindrical nozzles with a diameter of $18\,$mm. To improve isotropy and protect the particles from collisions with the pumps, a $13\,$mm square mesh is placed $3\,$cm downstream of the jets. Turbulence intensity is controlled through modulating the power supplied by pulse-width-modulation. For more information on the turbulence facility, the reader is referred to \cite{esteban_laboratory_2019}. The equipment is identical apart from the addition of the mesh and the power control system.
	
	Prior to releasing particles, the turbulence generated is characterised with particle image velocimetry (PIV). The flow was seeded with $56\,$\textmu m polyamide particles (Vestosint 2157). A laser sheet passing through the centre of the tank contained in the $xz$-plane is generated (Litron BERN 200-15PIV), and 3000 image pairs are taken at $0.8\,$Hz (VC-Imager Pro LX 16M). The interpulse time is set to $4000\,$\textmu s to limit the tracer displacements to approximately $6\,$px and reduce the out-of-plane displacements between image pairs.
	
\begin{figure}
		\centering
		\includegraphics[]{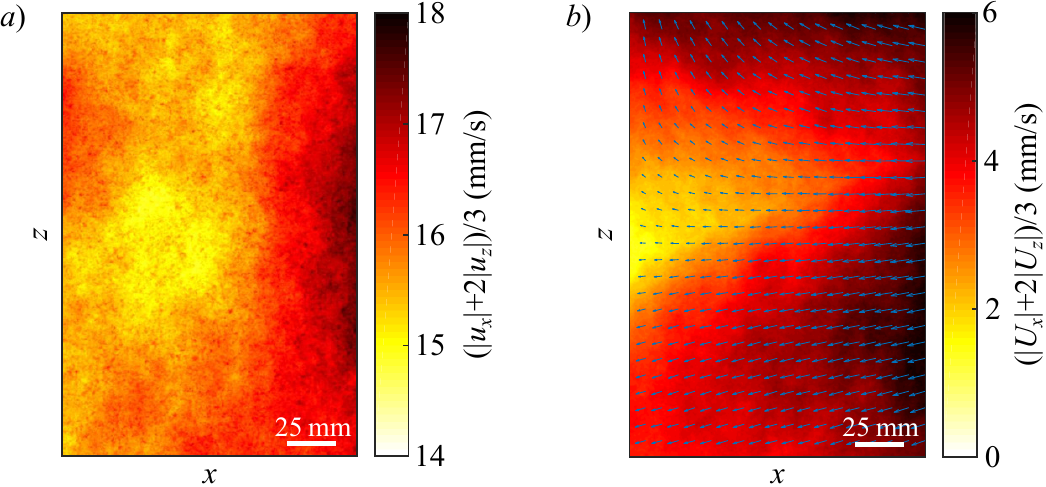}
		\caption{PIV measurements of the flow in the turbulence box. The time-averaged ({\it a}) root-mean-square flow velocity fluctuation field and ({\it b}) mean flow field at the middle of the tank. The subscripts $(x,z)$ denote the corresponding velocity components.}
		\label{Fig.TurbFlowField}
	\end{figure}

	To characterise the turbulence generated, we decompose the flow velocity into mean and fluctuating components $\boldsymbol{U}_f(\textbf{x},t) = \boldsymbol{U}_{mean}(\textbf{x},t) + \boldsymbol{u}_{fluc}(\textbf{x},t)$, where $\textbf{x}$ is the position vector. Figure \ref{Fig.TurbFlowField} shows the two fields, where $(U_x(\textbf{x}),U_z(\textbf{x}))$ and $(u_x(\textbf{x}),u_z(\textbf{x}))$ are the time-averaged $(x,z)$ components of $\boldsymbol{U}_{mean}$ and of the root-mean-square of $\boldsymbol{u}_{fluc}$ respectively. The fluctuations appear homogeneous although there is some horizontal mean flow caused by the synthetic jets emitted by the pumps. These are quantitatively expressed by the homogeneity deviation $HD$ and the mean flow factor $MFF$. Assuming symmetry about the $x$-axis \citep{carter_scale--scale_2017}, $u'_{rms} = (\overline{u_x} + 2\overline{u_z})/3$, where the line above denotes spatial averaging. Then $HD = 2\sigma_u/u'_{rms} = 0.07 \ll 1$ \citep{esteban_dynamics_2019}, where $\sigma_u$ is the standard deviation of $u'_{rms}$ in space. Thus the turbulence is indeed homogeneous. Denoting the mean flow speed by $U$, the relative magnitude of the mean flow is assessed by $MFF = U/u'_{rms} = 0.48$. While a small mean flow is present, velocity fluctuations still dominate so we believe it has no significant effect on the settling characteristics of the particles tested. Nonetheless, the global isotropy $\overline{u_x}/\overline{u_z} = 1.34 > 1$ shows the turbulence is mildly anisotropic. This implies the integral lengthscales and Taylor microscales depend on the direction of the velocity component and of the spatial separation.
	
\begin{figure}
		\centering
		\includegraphics[]{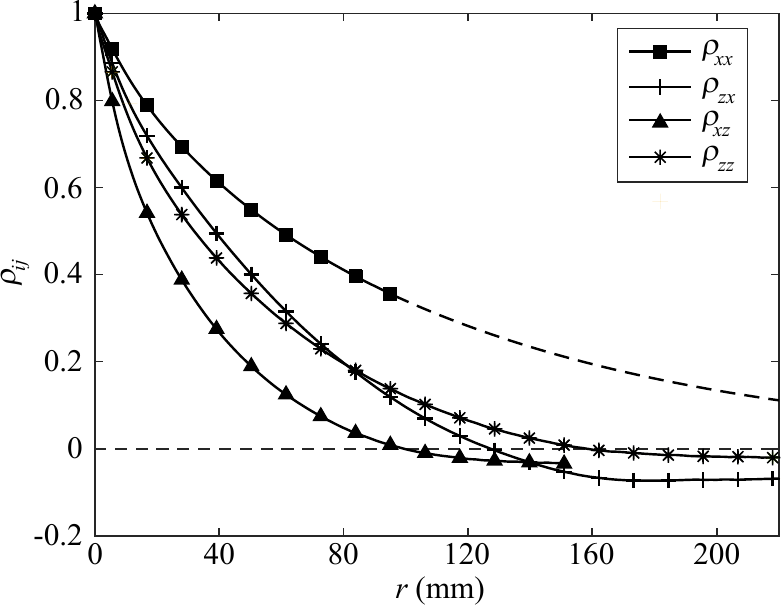}
		\caption{Autocorrelation functions $\rho_{ij}$ of the $j$-component velocity fluctuations along the $i$-direction. The solid lines give the measured data while the dashed line shows the exponential fit. The spacing between markers is not indicative of the resolution.}
		\label{Fig.Autocorrelation}
	\end{figure}

	Taking this into account, figure \ref{Fig.Autocorrelation} gives the various autocorrelation functions along the vertical and horizontal directions, $\rho_{ij}$. They decay as $r$ increases and approach 0 at $r \rightarrow +\infty$. Thus, we define the upper integration limit $r_0$ for the integral lengthscale $L_{ij}$ such that $\rho_{ij}(r_0)$ first reaches 0.01. This is in line with the suggestion in \cite{oneill_autocorrelation_2004}: taking $r_0$ as the first zero-crossing of $\rho_{ij}$ balances accuracy with ease of calculation. Furthermore, if the directly measured autocorrelation does not reach $\rho_{ij} \approx 0.01$, an exponential tail is fitted for $\rho_{ij} \leq 0.35$. Table \ref{Table.TurbStat} includes the various $L_{ij}$ found.
	
	Integral lengthscales involving velocity fluctuations along the $x$-direction are larger than those along $z$. This was also found by \cite{carter_scale--scale_2017} in a similar facility despite a different $Re_\lambda$, suggesting eddies were elongated by the larger fluctuations. Following their suggestion, the geometric mean of integral lengthscales involving fluctuations along one direction is taken to represent the size of the largest vortices in that orientation, i.e. $L_x = (L_{xx}L_{zx})^{\frac{1}{2}}$ for instance. To facilitate comparison with previous experiments, a conventional integral lengthscale assuming axisymmetry 
	\begin{equation}	\label{Eq.Axisymmetry}
	L_{turb} = \frac{L_x + 2L_z}{3}
	\end{equation}
	is evaluated too.
	
	The Taylor microscale along the $i$-direction of $j$-component velocity fluctuations, on the other hand, is evaluated according to its definition $\lambda_{ij} = \big(-\frac{1}{2}\frac{d^2\rho_{ij}}{dr^2}\big|_{r=0}\big)^{-0.5}$. To minimise PIV error, we only consider the first two values of $\rho_{ij}$ with a positive separation whose interrogation windows do not overlap \citep{adrian_particle_2011}. The horizontal intercept of the fitted parabola then equals $\lambda_{ij}$. The conventional longitudinal and transverse Taylor microscales $\lambda_f$ and $\lambda_g$ are found assuming axisymmetry in analogy to (\ref{Eq.Axisymmetry}).
	
	The related Reynolds number $Re_\lambda$ is also determined using the measured water temperature of $17\,^\circ$C. The direction-dependent values $Re_{\lambda,i} = \lambda_{g,i}\overline{u_i}/\nu$, where $\lambda_{g,i}$ is the transverse Taylor microscale involving $i$-component velocity fluctuations $\overline{u_i}$. The conventional axisymmetric $Re_\lambda$ and all the quantities discussed above are displayed in table \ref{Table.TurbStat}.
	
	All in all, these measurements show the background turbulence is homogeneous but mildly anisotropic.
	\bibliographystyle{jfm}
	\bibliography{SettlingParticles_vLBE}
\end{document}